\documentclass[sigconf]{acmart}
\usepackage{multicol}
\usepackage{multirow}
\usepackage{booktabs}
\usepackage{array}
\usepackage{lscape}
\usepackage{float}
\usepackage{colortbl}
\usepackage{geometry}

\definecolor{Goldenrod}{RGB}{85, 65, 13}
\definecolor{MyGray}{RGB}{194, 191, 190}

\newcolumntype{?}{!{\color{MyGray}\vrule width 0.3pt}}

\AtBeginDocument{%
  }

\setcopyright{acmlicensed}

\copyrightyear{2025}
\acmYear{2025}
\setcopyright{cc}
\setcctype{by}
\acmConference[CHI '25]{CHI Conference on Human Factors in Computing Systems}{April 26-May 1, 2025}{Yokohama, Japan}
\acmBooktitle{CHI Conference on Human Factors in Computing Systems (CHI '25), April 26-May 1, 2025, Yokohama, Japan}\acmDOI{10.1145/3706598.3713551}
\acmISBN{979-8-4007-1394-1/25/04}

\usepackage{xspace} 
\usepackage{xcolor}
\usepackage{hyperref}

\newcommand{\textsand}[1]{{\sffamily #1}}

\definecolor{appleredlight}{RGB}{255, 105, 97}
\definecolor{appleorangelight}{RGB}{255, 179, 64}
\definecolor{appleyellowlight}{RGB}{255, 212, 38}
\definecolor{applegreenlight}{RGB}{48, 219, 91}
\definecolor{applemintlight}{RGB}{102, 212, 207}
\definecolor{appleteallight}{RGB}{93, 230, 255}
\definecolor{applecyanlight}{RGB}{112, 215, 255}
\definecolor{applebluelight}{RGB}{64, 156, 255}
\definecolor{appleindigolight}{RGB}{125, 122, 255}
\definecolor{applepurplelight}{RGB}{218, 143, 255}
\definecolor{applepinklight}{RGB}{255, 100, 130}
\definecolor{applebrownlight}{RGB}{181, 148, 105}

\definecolor{applerednormal}{RGB}{255, 69, 58}
\definecolor{appleorangenormal}{RGB}{255, 159, 10}
\definecolor{appleyellownormal}{RGB}{255, 214, 10}
\definecolor{applegreennormal}{RGB}{48, 209, 88}
\definecolor{applemintnormal}{RGB}{99, 230, 226}
\definecolor{appletealnormal}{RGB}{64, 200, 224}
\definecolor{applecyannormal}{RGB}{100, 210, 255}
\definecolor{applebluenormal}{RGB}{10, 132, 255}
\definecolor{appleindigonormal}{RGB}{94, 92, 230}
\definecolor{applepurplenormal}{RGB}{191, 90, 242}
\definecolor{applepinknormal}{RGB}{255, 55, 95}
\definecolor{applebrownnormal}{RGB}{172, 142, 104}

\definecolor{applegrey}{RGB}{99, 99, 102}

\newcommand{\redn}[1]{\textcolor{applerednormal}{#1}}
\newcommand{\bluen}[1]{\textcolor{applebluenormal}{#1}}

\def\stage{\textsand{Stage}\xspace}
\def\stages{\textsand{Stages}\xspace}
\def\problem{\textsand{Problem}\xspace}
\def\problems{\textsand{Problems}\xspace}
\def\aim{\textsand{Aim}\xspace}
\def\aims{\textsand{Aims}\xspace}
\def\solution{\textsand{Solution}\xspace}
\def\solutions{\textsand{Solutions}\xspace}

\def\machine{\bluen{\textsand{Machine}}\xspace}
\def\machines{\bluen{\textsand{Machines}}\xspace}

\def\analyst{\redn{\textsand{Analyst}}\xspace}
\def\analysts{\redn{\textsand{Analysts}}\xspace}

\newcommand{\paran}[1]{(\hspace{1pt}#1{}\hspace{1pt})}

\newcommand{\revise}[1]{#1}

\usepackage{longfbox} 

\makeatletter
\newdimen\@tempdimd
\makeatother

\newfboxstyle{patternparamrad}{padding-top=2pt, padding-bottom=2pt, padding-left=4pt, padding-right=4pt, border-style=solid, height=6.5pt, border-radius=5pt}

\newfboxstyle{patternparam}{padding-top=2pt, padding-bottom=2pt, padding-left=3pt, padding-right=3pt, border-style=none, height=6.5pt}

\newcommand{\bluebox}[1]{\lfbox[patternparam, background-color=applebluenormal]{{\color{white}{{\normalfont \textsand{#1}}}}}}

\newcommand{\redbox}[1]{\lfbox[patternparam, background-color=applerednormal]{{\color{white}{{\normalfont \textsand{#1}}}}}}

\def\Preprocessing{\lfbox[patternparam, background-color=applebluenormal]{{\color{white}{{\normalfont \textsand{Preprocessing}}}}}\xspace}

\def\Preprocessingbf{\lfbox[patternparam, background-color=applebluenormal]{{\color{white}{{\normalfont \textsand{\textbf{Preprocessing}}}}}}\xspace}

\def\DR{\lfbox[patternparam, background-color=applebluenormal]{{\color{white}{{\normalfont \textsand{DR}}}}}\xspace}

\def\DimRedbf{\lfbox[patternparam, background-color=applebluenormal]{{\color{white}{{\normalfont \textsand{\textbf{Dimensionality Reduction}}}}}}\xspace}

\def\DRbf{\lfbox[patternparam, background-color=applebluenormal]{{\color{white}{{\normalfont \textsand{\textbf{DR}}}}}}\xspace}

\def\DimRed{\lfbox[patternparam, background-color=applebluenormal]{{\color{white}{{\normalfont \textsand{Dimensionality Reduction}}}}}\xspace}

\def\Evaluation{\lfbox[patternparam, background-color=applebluenormal]{{\color{white}{{\normalfont \textsand{Quantitative Evaluation}}}}}\xspace}

\def\Evaluationbf{\lfbox[patternparam, background-color=applebluenormal]{{\color{white}{{\normalfont \textsand{\textbf{Quantitative Evaluation}}}}}}\xspace}

\def\Visualization{\lfbox[patternparam, background-color=applebluenormal]{{\color{white}{{\normalfont \textsand{Visualization}}}}}\xspace}

\def\Visualizationbf{\lfbox[patternparam, background-color=applebluenormal]{{\color{white}{{\normalfont \textsand{\textbf{Visualization}}}}}}\xspace}

\def\Sensemaking{\lfbox[patternparam, background-color=applerednormal]{{\color{white}{{\normalfont \textsand{Sensemaking}}}}}\xspace}

\def\Sensemakingbf{\lfbox[patternparam, background-color=applerednormal]{{\color{white}{{\normalfont \textsand{\textbf{Sensemaking}}}}}}\xspace}

\def\Interaction{\lfbox[patternparam, background-color=applerednormal]{{\color{white}{{\normalfont \textsand{Interaction}}}}}\xspace}

\def\Interactionbf{\lfbox[patternparam, background-color=applerednormal]{{\color{white}{{\normalfont \textsand{\textbf{Interaction}}}}}}\xspace}

\def\Pioneer{\lfbox[patternparam, background-color=applepinknormal!20]{{\normalfont \textsand{Pioneer}}}\xspace}

\def\Pioneerbf{\lfbox[patternparam, background-color=applepinknormal!20]{{\normalfont \textsand{\textbf{Pioneer}}}}\xspace}

\def\Judge{\lfbox[patternparam, background-color=appleindigonormal!20]{{\normalfont \textsand{Judge}}}\xspace}

\def\Judgebf{\lfbox[patternparam, background-color=appleindigonormal!20]{{\normalfont \textsand{\textbf{Judge}}}}\xspace}

\def\Instructor{\lfbox[patternparam, background-color=applegreennormal!20]{{\normalfont \textsand{Instructor}}}\xspace}

\def\Instructorbf{\lfbox[patternparam, background-color=applegreennormal!20]{{\normalfont \textsand{\textbf{Instructor}}}}\xspace}

\def\Explorer{\lfbox[patternparam, background-color=applegrey!20]{{\normalfont \textsand{Explorer}}}\xspace}

\def\Explorerbf{\lfbox[patternparam, background-color=applegrey!20]{{\normalfont \textsand{\textbf{Explorer}}}}\xspace}

\def\Explainer{\lfbox[patternparam, background-color=appleyellownormal!20]{{\normalfont \textsand{Explainer}}}\xspace}

\def\Explainerbf{\lfbox[patternparam, background-color=appleyellownormal!20]{{\normalfont \textsand{\textbf{Explainer}}}}\xspace}

\def\Architect{\lfbox[patternparam, background-color=applebrownnormal!20]{{\normalfont \textsand{Architect}}}\xspace}

\def\Architectbf{\lfbox[patternparam, background-color=applebrownnormal!20]{{\normalfont \textsand{\textbf{Architect}}}}\xspace}

\newcommand{\redblock}{\raisebox{-0.22em}{\includegraphics[height=1.1em]{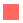}}\xspace}

\newcommand{\redtransblock}{\raisebox{-0.22em}{\includegraphics[height=1.1em]{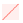}}\xspace}

\newcommand{\blueblock}{\raisebox{-0.22em}{\includegraphics[height=1.1em]{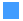}}\xspace}

\newcommand{\bluetransblock}{\raisebox{-0.22em}{\includegraphics[height=1.1em]{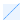}}\xspace}

\newcommand{\primaryicon}{\raisebox{-0.22em}{\includegraphics[height=1.1em]{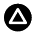}}\xspace}

\newcommand{\secondaryicon}{\raisebox{-0.22em}{\includegraphics[height=1.1em]{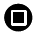}}\xspace}

\newcommand{\externalicon}{\raisebox{-0.22em}{\includegraphics[height=1.1em]{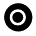}}\xspace}

\usepackage{changepage}

\renewcommand{\paragraph}[1]{
\vspace{4pt}
\noindent
\textbf{#1.}
}

\newcommand{\novparagraph}[1]{
\noindent
\textbf{#1.}
}

\usepackage{enumitem}

\setlist[itemize]{label=\raisebox{-0.13em}{\includegraphics[height=0.8em]{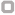}}, left=0.13in}

\begin{document}

\title[Unveiling High-dimensional Backstage]{Unveiling High-dimensional Backstage: A Survey for Reliable Visual Analytics with Dimensionality Reduction}


\author{Hyeon Jeon}
\affiliation{%
  \institution{Seoul National University}
  \city{Seoul}
  \country{Republic of Korea}}
\email{hj@hcil.snu.ac.kr}

\author{Hyunwook Lee}
\affiliation{%
  \institution{UNIST}
  \city{Ulsan}
  \country{Republic of Korea}
}
\email{hwlee0916@unist.ac.kr}

\author{Yun-Hsin Kuo}
\affiliation{%
 \institution{University of California, Davis}
 \city{Davis}
 \state{CA}
 \country{USA}}
\email{yskuo@ucdavis.edu}

\author{Taehyun Yang}
\affiliation{%
  \institution{Seoul National University}
  \city{Seoul}
  \country{Republic of Korea}}
\email{0705danny@snu.ac.kr}

\author{Daniel Archambault}
\affiliation{%
  \institution{Newcastle University}
  \city{Newcastle upon Tyne}
  \country{UK}
}
\email{Daniel.Archambault@newcastle.ac.uk}

\author{Sungahn Ko}
\affiliation{%
  \institution{UNIST}
  \city{Ulsan}
  \country{Republic of Korea}
}
\email{sako@unist.ac.kr}

\author{Takanori Fujiwara}
\affiliation{%
  \institution{Link\"oping University}
  \city{Link\"oping}
  \country{Sweden}}
\email{takanori.fujiwara@liu.se}

\author{Kwan-Liu Ma}
\affiliation{%
 \institution{University of California, Davis}
 \city{Davis}
 \state{CA}
 \country{USA}}
\email{klma@ucdavis.edu}

\author{Jinwook Seo}
\authornote{Corresponding Author}
\affiliation{%
  \institution{Seoul National University}
  \city{Seoul}
  \country{Republic of Korea}}
\email{jseo@snu.ac.kr}

\renewcommand{\shortauthors}{}

\begin{abstract}
  Dimensionality reduction (DR) techniques are essential for visually analyzing high-dimensional data. 
However, visual analytics using DR often face \textit{unreliability}, stemming from factors such as inherent distortions in DR projections.
This unreliability can lead to analytic insights that misrepresent the underlying data, potentially resulting in misguided decisions.
To tackle these reliability challenges, we review 133 papers that address the unreliability of visual analytics using DR. 
Through this review, we contribute (1) a workflow model that describes the interaction between analysts and machines in visual analytics using DR, and (2) a taxonomy that identifies where and why reliability issues arise within the workflow, along with existing solutions for addressing them.
Our review reveals ongoing challenges in the field, whose significance and urgency are validated by five expert researchers.
This review also finds that the current research landscape is skewed toward developing new DR techniques rather than their interpretation or evaluation, where we discuss how the HCI community can contribute to broadening this focus.


\end{abstract}



\received{20 February 2007}
\received[revised]{12 March 2009}
\received[accepted]{5 June 2009}

\maketitle

\section{Introduction}

\label{sec:intro}

Dimensionality reduction (DR) serves as a backbone of data visualization and visual analytics \cite{wenskovitch18tvcg, kwon18tvcg, nonato19tvcg, bruneau15neurocomputing}. 
DR techniques such as $t$-SNE \cite{maaten08jmlr}, UMAP \cite{mcinnes2020arxiv}, and RadViz \cite{rubio-sánchez16tvcg} generate low-dimensional representations of high-dimensional data while preserving key structural characteristics of the original data, including local neighborhood structure and global cluster arrangement. 
DR projections are widely incorporated in many visual analytics systems and applications \cite{kwon18tvcg, narechania22tvcg, cheng23tvcg} across diverse domains such as bioinformatics \cite{cheng23tvcg, becht19nature, narayan21nature, moon2019visualizing, amir2013visne}, natural language processing \cite{li16arxiv, templeton24tct}, and human-computer interaction \cite{bai21arxiv, lim23chi}. DR has also been used to explain machine learning models \cite{rauber17tvcg, yeh24tvcg}, including cutting-edge foundation models like large language models \cite{templeton24tct}.

Still, visual analytics leveraging DR can easily become unreliable: derived insights or knowledge from visual analytics may not accurately reflect the underlying data, potentially leading to flawed decisions.
For example, because low-dimensional spaces have a fewer degree of freedom than high-dimensional spaces, distortions are inevitable in DR projections. 
These distortions can reduce the accuracy of the projections~\cite{aupetit07neurocomp, jeon21tvcg, faithfulness}, misrepresenting the original data and potentially leading analysts to incorrect conclusions about its structure.
Such erroneous insights or decisions can cascade downstream, negatively affecting subsequent analyses or research that relies on them.
To address these challenges, the visual analytics community has thus conducted numerous studies to enhance the reliability of visual analytics using DR.

However, we need more comprehensive guidance to help analysts address the full scope of unreliability when using DR. 
Existing surveys \cite{nonato19tvcg, espadoto21tvcg, sorzano14arxiv} focus on guiding analysts in selecting DR techniques that produce accurate projections for specific data or tasks. For example, they recommend using $t$-SNE or UMAP to reliably support neighborhood investigation tasks. While these surveys contribute to making visual analytics more reliable by increasing practitioners' awareness of the design and purpose of DR techniques, unreliability can still arise for various other reasons, such as suboptimal hyperparameter settings \cite{kobak21nb, xiang21fig} or the inherent instability of certain DR techniques \cite{mcinnes2020arxiv, kobak21nb}.
Other approaches beyond DR technique selection, including making DR more interactive \cite{appleby22cgf, fujiwara21tvcg} or interpretable \cite{lespinats07tnn, jeon21tvcg, bian21tvcg}, are proposed to remedy these problems.


Our research aims to coordinate these diverse efforts,  thereby contributing to achieving more reliable visual analytics with DR. 
We first carefully review the literature, identifying 133 relevant papers from diverse research fields, including data visualization, human-computer interaction, and machine learning. 
Based on this systematic review, we develop a  \textbf{workflow model} that outlines the interaction between analysts and machines when conducting visual analytics with DR.
 Grounded in the identified papers and previous theories in visual analytics, this model enhances our understanding of how analysts use DR in practice. It offers \revise{a comprehensive explanation of DR usage}, covering the entire analytic process---from preprocessing to visualizations and sensemaking.
Finally, we organize the reviewed papers into a \textbf{taxonomy} structured around four dimensions: \stage, \problem, \aim, and \solution, each branch mapped to a component of the workflow model or defined by those components. 
A meta-analysis reveals six prevalent clusters of papers, offering deeper insights into the overall research landscape.



Our analysis identifies several \textbf{ongoing challenges}, 
 including the limited discussion extending beyond 2D scatterplots and the lack of available libraries to support more advanced DR techniques.
Eight expert researchers in the relevant fields validated the significance and urgency of these challenges.
\revise{
Furthermore, our survey reveals an imbalance in the research landscape, with many papers proposing new techniques but few focusing on their evaluation or interpretation. 
To fill this gap, we \textbf{call for the HCI community} to invest more efforts to enhance the interpretability of DR techniques and make it easier to use DR reliably. As a seminal step, we develop and release an actionable guide to help researchers navigate the literature included in our survey.
}

In summary, this research makes four key contributions: 
\begin{itemize}
    \item We present a \textbf{workflow model} that comprehensively details the visual analytics procedure using DR.
    \item We propose a \textbf{taxonomy} that categorizes 133 relevant studies in the field.
    \item We identify three significant and urgent \textbf{ongoing challenges} in DR-driven visual analytics. 
    \item \revise{We reveal an imbalance in the relevant literature, where we \textbf{prompt the HCI community} to mitigate it.}

\end{itemize}
To the best of our knowledge, this is the first survey specifically addressing the issue of unreliability in visual analytics for DR techniques.
We thus hope this research will serve as a valuable reference not only for practical DR applications but also for future investigations into reliable visual analytics. We provide an interactive browser of this survey at \url{https://dr-reliability.github.io/demo}.

\section{Related Work}

We discuss three relevant areas of previous work: 
(1) existing surveys on DR techniques and evaluation metrics, (2) investigations on the practical usage of DR in visual analytics, and (3) theoretical models for visual analytics.

\subsection{Surveys on Dimensionality Reduction}

We review existing surveys on DR techniques and evaluation metrics. These surveys share a common goal with our research: enhancing the reliability of visual analytics using DR.
\revise{
Our research differentiates itself by addressing a broader spectrum of threats that can compromise reliability, spanning various stages of the analytical workflow beyond selecting DR techniques and metrics. 
}

\novparagraph{Surveys on DR techniques}
These surveys aim to reveal the pros and cons of existing DR techniques, providing guidance for practitioners in selecting the most appropriate technique that matches the analysts' intentions, e.g., target tasks \cite{nonato19tvcg, vandermaaten09jmlr, sorzano14arxiv, cunningham15jmlr, fodor02survey, yin07ijac, engel12irtg}.
However, these literature reviews often lack quantitative validation of the performance of DR techniques, e.g., accuracy in preserving the structure of high-dimensional datasets, limiting the extent to which their guidance can be applied. For example, a technique may work appropriately for dense datasets but poorly for sparse datasets \cite{espadoto21tvcg}.
Researchers thus contribute benchmark studies that incorporate the measurements and analyses of projection accuracy or scalability \cite{espadoto21tvcg, bunte12neco, vandermaaten09jmlr, atzberger24tvcg} to fill this gap. Recently, Espadoto et al. \cite{espadoto21tvcg} conduct a large-scale study that shows how the performance of DR techniques is affected by the dataset traits (e.g., dimensionality, number of data points) or hyperparameter settings. Xia et al. \cite{xia22tvcg} and Etemadpour et al. \cite{etemadpour15tvcg} also contribute large-scale benchmark studies, where they directly examine human task performance by conducting user studies.

\paragraph{Surveys on evaluation metrics}
Although DR techniques are specialized to preserve certain aspects of the original data, they inherently cannot escape from distortions \cite{nonato19tvcg, colange19vis, jeon21tvcg, espadoto21tvcg, lespinats07tnn}.
Thus, many quality metrics are proposed to assess the accuracy of DR projections in preserving structural characteristics of the original high-dimensional data \cite{venna06nn, jeon21tvcg, jeon24tvcg, colange20neurips}. As with DR techniques, suitable use of quality metrics is essential in conducting reliable data analysis. For example, to identify DR projections that reliably support cluster analysis, metrics specialized to cluster structure (e.g., Steadiness \& Cohesiveness \cite{jeon21tvcg}) are recommended to test the projections.

Reviews on DR quality metrics thus contribute as guidelines for selecting proper metrics. For example, Nonato and Aupetit \cite{nonato19tvcg} and Thurn et al. \cite{thrun23make} taxonomize existing quality metrics based on their target structural characteristics (e.g., local neighborhood or global pairwise distance). 
Lee and Verleysen \cite{lee09neurocomp} provide an extensive review of the quality metrics that assess the preservation of neighborhood structure. 
A recent study from Jeon et al. \cite{jeon23vis} not only conducts the review but deploys the surveyed metrics as a unified Python library.

\paragraph{\textit{Our contribution}}
Existing literature reviews and benchmarks enhance the unreliability of visual analytics by addressing a lack of awareness about the disadvantages and advantages of DR techniques and quality metrics in supporting diverse analytic tasks.
This is done by providing guidelines in selecting DR techniques and metrics that \revise{optimally aligns with} the analysts' intention (e.g., task, prior domain knowledge) and data.
In this work, we demonstrate that \revise{factors beyond optimality} can compromise the reliability of visual analytics \revise{(e.g., interpretability or stability of DR techniques)}.
\revise{We also show that the efforts for addressing this unreliability go beyond the selection of proper techniques or quality metrics (\autoref{sec:taxonomy}).}
\revise{For example, our survey shows that visual analytics systems that explain DR techniques \cite{cutura18esann} or unified libraries \cite{jeon23vis} can even contribute to enhancing reliability.}
Doing so offers a more comprehensive taxonomy and actionable guidelines to help analysts conduct more reliable visual analytics using DR.

\subsection{Investigation on the Practical Usage of Dimensionality Reduction}

Given the wide usage of DR, several works investigate how DR is used for visual analytics \cite{sacha17tvcg, brehmer14beliv,espadoto20visgap}. 
Brehmer et al. \cite{brehmer14beliv} conduct interviews with data analysts, revealing a workflow of using DR for analyzing cluster structure or interpreting the meaning of dimensions. 
Based on task sequences, they derive guidelines for a more comprehensive evaluation of DR-based visualizations. 
Sacha et al. \cite{sacha17tvcg} conduct a literature review to identify how analysts interact with DR projections.
They contribute a framework systematizing human-DR interaction and then use it to understand and compare visual analytics systems using DR. 
They also use the proposed framework to suggest research directions for enriching interaction with DR.
Espadoto et al. \cite{espadoto20visgap} explore workflows for developing, evaluating, and deploying DR techniques for practical applications. They further identify the toolkits or libraries available for each workflow step. Furthermore, they discuss the challenges of using such toolkits in practice, e.g., the lack of benchmark datasets to evaluate DR techniques. 

\paragraph{\textit{Our contribution}}
Previous research focuses on specific scenarios to offer insights into using DR in visual analytics.
In contrast, we offer a more comprehensive explanation of DR usage, covering the entire workflow from data preprocessing to visualization and sensemaking. This approach gives a broader understanding of DR usage in practice.
\revise{Grounded in this workflow, our study comprehensively captures} reliability problems \revise{in DR-based visual analytics} that previous studies have not fully explored.

\subsection{Theoretical Models for Visual Analytics Workflow}

\label{sec:relmodel}

Visual analytics and data mining communities have proposed diverse models that explain how analytic procedures derive knowledge from data. 
Van Wijk \cite{wijk06tvcg} propose a model of how each component in visual analytics (e.g., perception, interaction, visualization specification) affects knowledge. Van Wijk's model has been further complemented by Green et al. \cite{green09infovis}. Fayyad et al. \cite{fayyad96aaai} propose the
Knowledge Discovery in Databases (KDD) process, which describes the data mining stages that raw data should go through to generate knowledge.
Chen and J\"aenicke \cite{chen10tvcg} explain the visual analytics process by aligning it with information theory. 
Keim et al. \cite{keim08ivs} review existing theories and models to comprehensively explain the visual analytics process. 
Based on the previous workflow models, Sacha et al. explain the role of visual analytics in generating knowledge \cite{sacha14tvcg}.

\paragraph{\textit{Our contribution}}
This research introduces a theoretical workflow model
describing how analysts and machines behave when conducting visual analytics (\autoref{sec:workflow}).
\revise{Our model uses these previous workflows as the theoretical backbone.
For example, our model follows the global structure proposed by Van Wijk \cite{wijk06tvcg}. }
\revise{Nevertheless, our model examines each component of the structure in detail, providing a fine-grained explanation of how DR-based visual analytics takes place.} 
Moreover, we not only contribute a workflow model but also discuss the unreliability that can arise within the workflow, providing practical guidance to help analysts make their visual analytics reliable.

\section{Protocol}

\begin{figure*}
    \centering
    \includegraphics[width=\textwidth]{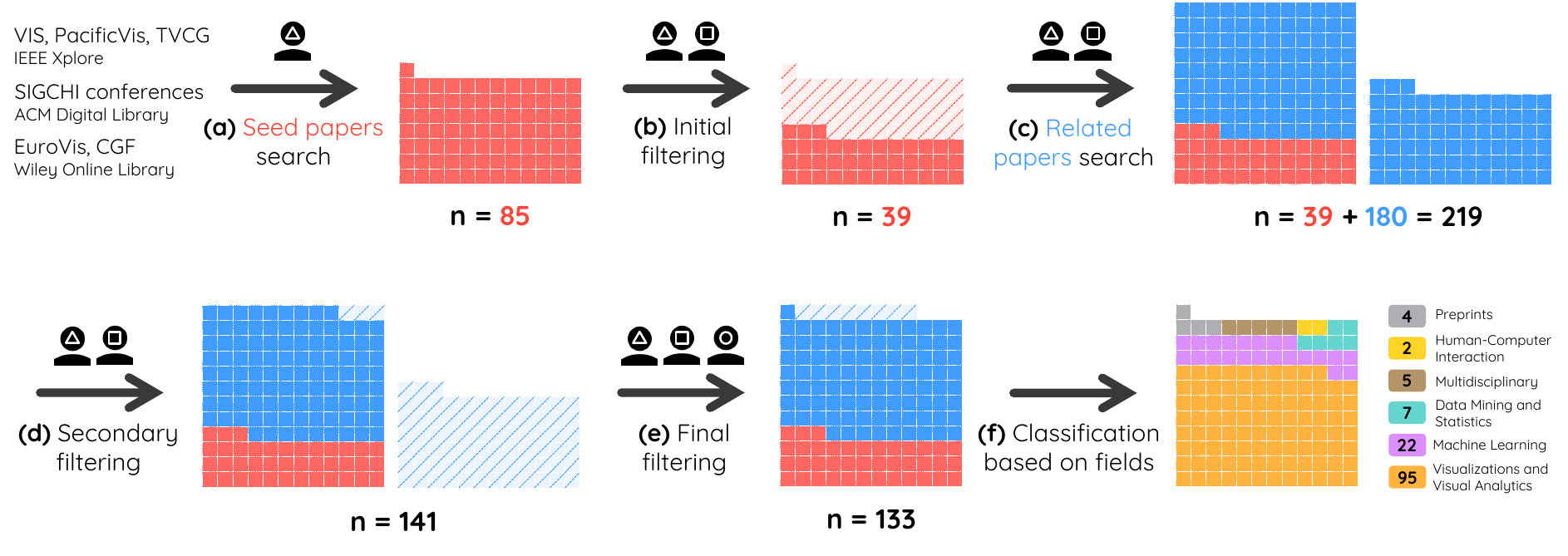}
    \caption{\textit{Procedure of selecting papers (a-e) and their classification based on research fields (f).}
    (a) We first search seed papers published in major visualization and human-computer interaction venues (\redblock), and (b) filter out papers that do not fall within our survey scope (\redtransblock). (c) We then extend our paper collection (\blueblock) by screening the related works and backgrounds of the seed papers. (d, e) Finally, by extensively reviewing and filtering out unrelated papers (\bluetransblock), we finalize a total of 133 papers. (f) We classify the papers based on their fields. The icons above each arrow represent the authors (\primaryicon, \secondaryicon, and \externalicon) that are involved in the corresponding step. 
    }
    \Description{This figure outlines the procedure for selecting papers (a-e) and their classification based on research fields (f). The process is illustrated with a series of steps: (a) Seed papers search: Seed papers (represented in red blocks) are selected from major visualization and human-computer interaction venues, such as VIS, PacificVis, TVCG, SIGCHI conferences, EuroVis, and others. (b) Initial filtering: Papers that do not fit within the survey scope are filtered out (shown as red shaded blocks), reducing the paper count from 85 to 39. (c) Related papers search: Related works from the seed papers' references and backgrounds are collected (depicted in blue blocks), bringing the total paper count to 219 (39 seed papers + 180 related papers). (d) Secondary filtering: Further filtering removes additional unrelated papers (shown as blue shaded blocks). (e) Final filtering: The paper count is refined to a total of 133 after reviewing the related papers thoroughly. (f) Classification: The papers are classified based on their research fields, visualized as colored blocks with labels indicating specific domains such as Preprints (4), Human-Computer Interaction (2), Multidisciplinary (5), Data Mining and Statistics (7), Machine Learning (22), and Visualizations and Visual Analytics (95). The icons above the arrows indicate the author involvement at each stage of the process.}
    \label{fig:procedure}
\end{figure*}

We discuss our search strategy for identifying relevant papers and provide an overview of the papers we select.
We then explain in detail how we develop our workflow model and taxonomy. 

\subsection{Paper Selection}

\label{sec:selectpaper}

We outline our survey scope and the procedure for selecting papers. 
We also provide examples of papers included and excluded in our survey. Furthermore, we show how the papers are distributed across the years and fields, confirming the wide coverage of our survey.

\paragraph{Survey scope}
We search papers that address unreliability from visual analytics using DR. 
The survey does not focus on unreliability primarily related to data-agnostic visualizations (e.g., overplotting in scatterplots) or other machine learning techniques (e.g., clustering or regression). The example papers for inclusion and exclusion are as follows:

\textit{Examples for inclusion:}
\begin{itemize}
    \item Studies that improve the accuracy of existing DR technique, e.g., by modifying loss functions or distance functions \cite{vu21ijcnn, tenenbaum00science, jeon22vis, narayan21nature}. These studies support analysts in conducting their tasks more accurately.
    \item Studies that propose new DR quality metrics \cite{jeon21tvcg, motta15neurocomp, sips09cgf} or improving the reliability of existing metrics \cite{jeon24tvcg, lee10prl}. These studies help practitioners identify DR projections optimal for their tasks based on metric scores.  
    \item Studies that propose visualizations or interaction techniques that explain distortions in projections \cite{jeon21tvcg, lespinats07tnn, colange19vis, aupetit07neurocomp}. These studies contribute to making practitioners more aware of the existence of inaccuracy of projections due to distortions.
    \item Benchmark studies that provide an extensive comparison of existing DR techniques \cite{espadoto21tvcg, xia22tvcg, etemadpour15tvcg, vandermaaten09jmlr}. These benchmarks help practitioners become more aware of the pros and cons of existing techniques and select the most appropriate techniques for given datasets and tasks.
\end{itemize}

\textit{Examples for exclusion:}
\begin{itemize}
    \item Studies that deal with reliability problems for general visualizations and thus are not always relevant to visual analytics using DR, e.g., visual perception of scatterplots \cite{abbas19cgf, jeon24tvcg2}. 
    \item Studies that mainly focus on improving the computational efficiency of existing DR techniques \cite{fu19kdd, kim17aaai}.
    \item Studies that aim to remedy reliability problems related to other types of ML techniques, e.g., clustering \cite{lewis12amcss, davidson01infvis}, or visualizations, e.g., Parallel Coordinates Plot, glyphs, that are used to analyze high-dimensional data \cite{yang03vissym, ross04cmv}.
\end{itemize}


\paragraph{Procedure}
Our procedure of selecting papers is inspired by existing surveys in the data visualization field \cite{junpeng24tvcg, shin23tvcg, qaudri22tvcg}.
Three authors search the papers in May 2024.

\textit{First, we search seed papers and filter irrelevant papers.}
A primary author search papers published in the major visualization and HCI venues---VIS (InfoVis, VAST, and SciVis), TVCG, PacificVis, EuroVis, CGF, and SIGCHI affiliated conferences---using the corresponding digital libraries: IEEE Xplore, Wiley Online Library, and ACM Digital library (\autoref{fig:procedure}a).  
We focus on visualization and HCI venues as our primary focus is to investigate the use of DR in visual analytics. 


The primary author query (\textit{``Dimensionality Reduction''} \textsc{OR} \textit{``Dimension Reduction''} OR \textit{``Multidimensional Projection''} \textsc{OR}  \textit{``Multidimensional Scaling''}) \textsc{AND} (\textit{``Reliability''} \textsc{OR} \textit{``Trustworthy''} \textsc{OR} \textit{``Distortion''} \textsc{OR} \textit{``Uncertainty''} \textsc{OR} \textit{``Quality''}) for paper title and abstract. 
We do not set specific time limits to make our list of papers more comprehensive.
As a result, we identify 85 papers in total (\autoref{fig:procedure}a; \redblock). Then, 
two authors screen the title, abstract, and introduction of the papers and filter out papers that are out of our scope (\autoref{fig:procedure}b; \redtransblock).
We filter out papers only if two authors agree to reduce the risk of excluding relevant papers. 
The procedure ends up with 39 seed papers.

\textit{Second, we extend the search space to the related works of seed papers. }
The two authors conduct the second screening of the seed papers to curate a list of papers that likely are within our survey scope.
We review the introduction section if a paper does not have such sections. Note that a paper is included in our collection if at least one of two authors thinks that the paper is relevant to our study to minimize the risk of overlooking relevant papers. As a result, we additionally find 180 papers (\autoref{fig:procedure}c; \blueblock).
However, as this extension is only done by reviewing the contents of seed papers, two authors again screen the title, abstract, and introduction of these new papers and filter out irrelevant papers (\autoref{fig:procedure}d; \bluetransblock), where we conclude with 141 papers (including 39 seed papers) in total. 

\textit{At last, we conduct the final screening and filtering of papers.}
We review the title, abstract, introduction, and method sections to ensure that our collection contains no false positives. An additional author is involved in this step to bring a fresh perspective to the screening.
We keep papers only if all three authors agree to include the papers. We exclude eight papers, resulting in a final collection of \textbf{133} papers (\autoref{fig:procedure}e).


\begin{figure}
    \centering
    \includegraphics[width=\linewidth]{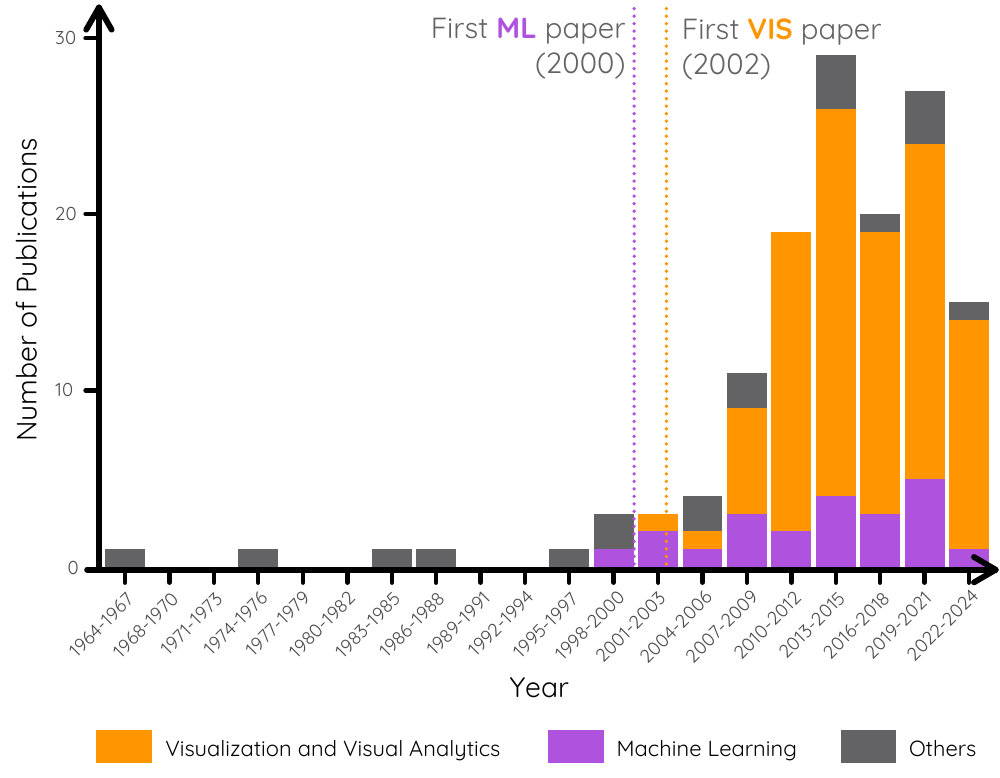}
    \caption{
    \textit{The distribution of collected papers over the years.}
    The papers we identify are dominantly incorporated in \textcolor{applepurplenormal}{machine learning (ML)} and \textcolor{appleorangenormal}{visualization (VIS)} fields.
    The number of published papers dramatically increases around the early 2000s. 
    }
    \Description{The bar chart (histogram) depicts the number of publications per year from 1964 to 2024 across three categories: "Visualization and Visual Analytics" (orange), "Machine Learning" (purple), and "Others" (gray). The y-axis, labeled "Number of Publications," ranges from 0 to 30, while the x-axis shows intervals of 3 to 4 years. Notable events are marked by vertical dashed lines: the first "Machine Learning" paper in 2000 (purple) and the first "Visualization and Visual Analytics" paper in 2002 (orange). Before 2000, the number of publications was minimal, but starting around 2003, there was a gradual increase, followed by a noticeable surge around 2012, with publications peaking between 2015 and 2018. The "Visualization and Visual Analytics" category dominates the later years, while "Machine Learning" and "Others" show smaller but consistent contributions. The legend at the bottom clarifies the color coding for each category.}
    \label{fig:time_field_analysis}
\end{figure}

\paragraph{Metadata analysis}
We want to identify the trends in relevant research over different periods and research fields. 
For this purpose, we classify the collected papers according to the research fields of the journals, conferences, and workshops in which they were published (\autoref{fig:procedure}f). We do so by reviewing the titles and descriptions provided by the venues.
We then investigate how the total number of papers and the number of papers in each field have changed over the years.

Our classification divides the papers into six groups: \textit{Visualization and visual analytics}, \textit{Machine learning}, \textit{Data mining and statistics}, \textit{Multidisciplinary}, \textit{Human-computer interaction}, and \textit{Preprints} (\autoref{fig:procedure} right bottom). 
The diversity of fields indicates that the reliability of data visualizations using DR is crucial in many research areas.
This implies that our survey is likely to impact a variety of domains and research fields.
The result is unsurprising since DR is widely employed to analyze high-dimensional data and present findings in various fields \cite{cheng23tvcg, li16arxiv, templeton24tct, bai21arxiv, lim23chi} (\autoref{sec:intro}). 

We also find that relevant papers have increased significantly since early 2000 (\autoref{fig:time_field_analysis}). 
This rise indicates that our topic is gaining more importance in these fields. 
Overall, our metadata analysis confirms the need for comprehensive knowledge of how the reliability of visual analytics using DR can be improved.


\subsection{Workflow Model and Taxonomy Design}

We establish our workflow model (\autoref{sec:workflow}) and taxonomy (\autoref{sec:taxonomy}) by conducting the iterative thematic analysis.
Three authors involve the process as coders. The detailed process is as follows.

\paragraph{Initial review and discussion}
Three coders individually read 133 collected papers and write a description describing when and why unreliability occurs in visual analytics and how the papers address such a problem.
In this stage, the coders note their findings in free-form texts to add depth to the descriptions.

\paragraph{Workflow model design}
Based on the descriptions, three coders design the workflow model through iterative discussions.
The coders first establish three main considerations in the design process through iterative discussions. 
\begin{itemize}
    \item \textbf{(C1)} The model should clearly distinguish the role of analysts and machines (i.e., computers).
    \item \textbf{(C2)} The model should cover all analytic procedures (e.g., preprocessing, execution of DR, evaluation of projections). 
    \item \textbf{(C3)} The model should correspond to or translate into existing models explaining visual analytics workflow.
\end{itemize}
We set C1 to clearly distinguish the role of humans (analysts) or computers (machines) in conducting visual analytics using DR. C2 is established to ensure that the workflow and the taxonomy (\autoref{sec:taxonomy}) can comprehensively cover the reliability problems stemming from all steps of the analytic workflow. 
Lastly, C3 is set to anchor our model in well-established and validated previous research,reducing its vulnerability to unexpected errors \revise{-- a common approach in the visualization field for building new theoretical models \cite{sacha14tvcg, qianwen23tvcg, green08vast, behrisch18cgf}}.

Three coders collaboratively revise the model after one coder drafts the model based on these considerations. The coders conduct four meetings to reach an agreement, where three more authors occasionally participate in this discussion. 
Please refer to \autoref{sec:workflow} for a detailed model description.

\paragraph{Taxonomy design and paper classification}
Based on the initial description of papers and workflow model, three coders iteratively design the taxonomy and classify papers. At first, three coders draft the taxonomy together. The main consideration in this step is the alignment of the workflow model and the taxonomy; every branch within the taxonomy should be a component of the workflow or explained by the components. 

Using our initial taxonomy, three coders independently classify the papers.
The initial agreement measured by Cohen's $\kappa$ is $0.63$ on average. 
Three coders then iteratively revise the taxonomy and classification results until they reach an agreement. After two iterations, the final taxonomy and classification are established, which we detail in \autoref{sec:taxonomy}.

\section{Workflow Model}

\label{sec:workflow}

\begin{figure*}
    \centering
    \includegraphics[width=\linewidth]{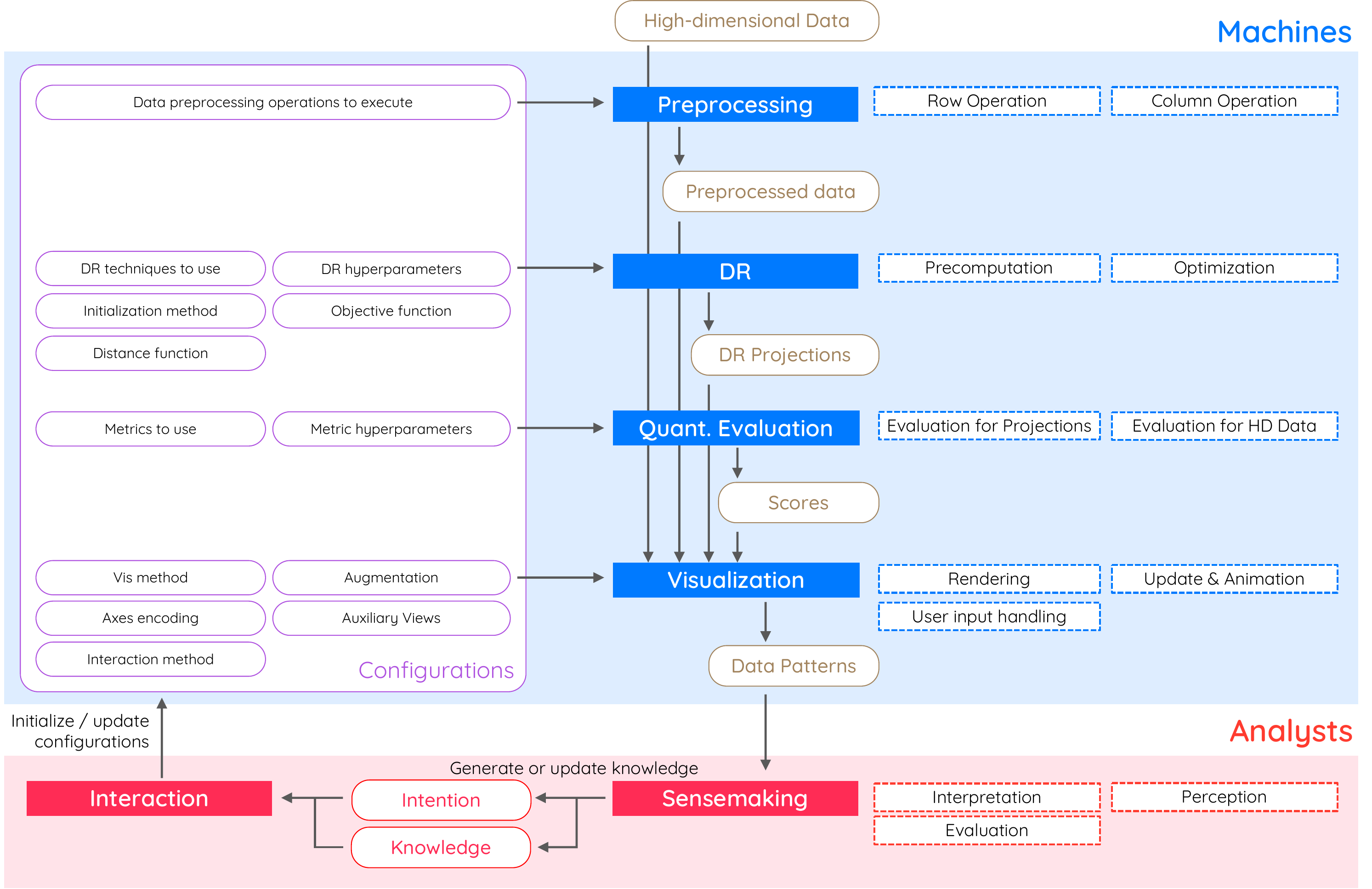}
    \caption{\textit{The illustration of our workflow model.} The model explains how an \analyst and a \machine interact while conducting visual analytics using DR. 
    Each stage of visual analytics executed by analysts and machines is represented by \redbox{red} and \bluebox{blue} rectangles, respectively, and the input and output of each stage are designated by arrows.  }
    \Description{This figure illustrates the workflow model for visual analytics using dimensionality reduction (DR), depicting interactions between analysts and machines. The workflow is divided into two main sections: Machines (blue) and Analysts (red), with distinct stages for each side. Machines Section: Preprocessing: High-dimensional data undergoes various preprocessing operations (row and column operations) to prepare the data for analysis. Dimensionality Reduction (DR): The preprocessed data is fed into DR techniques, where parameters such as hyperparameters, initialization methods, objective functions, and distance functions are configured. The DR step generates projections of the high-dimensional data. Quantitative Evaluation: The DR projections are evaluated using quantitative metrics, generating scores for projections and high-dimensional data evaluation. Visualization: The evaluated data patterns are visualized through rendering, updates, animations, and user input handling. Analysts Section: Interaction: Analysts interact with the system, updating configurations and feeding their intention and knowledge into the process. Sensemaking: Through sensemaking, analysts generate or update their knowledge, involving interpretation, perception, and evaluation of the visualized data patterns. Configurations: A separate section highlights the various configurations that can be updated by the analysts, such as DR techniques, metrics, visualization methods, and interaction methods. The stages are connected by arrows showing the flow of input and output between the machine and analyst roles, creating a continuous cycle of analysis and refinement.}
    \label{fig:workflow}
\end{figure*}

Grounded by the papers we identified (\autoref{tab:taxonomy1}) and previous theories on visual analytics (\autoref{sec:relmodel}), we design a workflow model explaining how DR is used in visual analytics. 
The workflow model enhances our understanding of how analysts conduct visual analytics using DR. 
In this study, the workflow contributes to building a more comprehensive taxonomy of the papers we identify (\autoref{sec:taxonomy}).

\sloppy
Our workflow model (\autoref{fig:workflow}) consists of two main actors (\analysts and \machines) and six stages \paran{\Preprocessing, \DR or \DimRed, \Evaluation, \Visualization, \Sensemaking, and \Interaction}. 
The first four stages are carried out by \machines, and the last two stages are executed by \analysts. 
Inspired by van Wijk's visualization model \cite{wijk06tvcg}, we describe visual analytics using DR as an iterative, looped process where analysts continually adjust configurations to update visualizations based on newly acquired knowledge.
Our model also reflects a common workflow of using machine learning techniques through iterative monitoring and refinement of specifications \cite{kapoor10chi, endert17cgf}.
We further suggest that high-dimensional data cascades from previous stages to subsequent stages executed by \machines, with each stage enriching the data, for example, by generating DR projections \paran{\DR} or evaluating their accuracy \paran{\Evaluation}.

\fussy

In the following sections, we first detail the labor of \analysts and \machines (\autoref{sec:analyst}). We then describe each stage of the analytics workflow (\autoref{sec:stages}).

\subsection{Analysts and Machines}

\label{sec:analyst}
Visual analytics cannot be established without \revise{both data \analysts and \machines.}
analysts initialize and update configurations that capture both their intentions (e.g., task, requirements, hypothesis) and knowledge, which is subsequently fed into \machines to guide their behavior.

\paragraph{\analysts}
As demonstrated in the knowledge generation model proposed by Sacha et al. \cite{sacha14tvcg}, analysts' main goal in our workflow is to perform \Sensemaking, thereby generating knowledge that is useful and also reliably reflects the original data. \revise{Detailed descriptions of analysts' actions are as follows.}

\revise{\textit{At first, analysts initialize configurations.}}
 This is done by interacting with \machines, e.g., by manipulating visualizations or writing a program (\Interaction stage). 
The setup is done based on analysts' knowledge (e.g., domain knowledge about the target data) and intention.
For example, analysts may configure machines to execute DR techniques that focus on local structure (e.g., $t$-SNE or UMAP) to perform local neighborhood investigation.
Similarly, when analysts' domain knowledge informs that certain attributes have low significance in the analysis, they can reduce the weight of those attributes in the distance function. Operational knowledge, encompassing the analysts' expertise in configurations, serves as a ``mental toolbox'' to support this process. Without prior knowledge of the configurations that fit their needs, analysts will have difficulty in establishing appropriate configurations. 
For example, analysts may hardly use $t$-SNE or UMAP for tasks related to the local structure if they do not know that such techniques fit the task well. 

\revise{\textit{Analysts then generate or update knowledge based on output visualizations.}}
After analysts deliver the configurations, the machines process the input high-dimensional data and visually convey data patterns to the analysts. Then, analysts perform \Sensemaking, e.g., perceiving and interpreting data patterns, thereby updating knowledge and intention. 
Please refer to Sacha et al. \cite{sacha14tvcg} or Kim et al. \cite{kim21tvcg, kim19chi} for a general explanation of the sensemaking procedure in visual analytics. 


\revise{\textit{Analysts again interact with machines to update configurations based on the updated knowledge and intention}}. This happens when analysts think the current configurations and resulting data patterns can hardly validate new hypotheses. For example, analysts may newly want to explore the detailed attribute values of clusters they discovered in the current DR projection. In such a case, the analysts need to reconfigure the \Visualization, e.g., by adding auxiliary views like parallel coordinate plots that can better depict the detailed attribute values of the clusters.

\sloppy
\paragraph{\machines}
In our workflow model, machines process the input high-dimensional data based on the given configurations. 
\revise{
It comprises four stages: \Preprocessing, \DimRed, \Evaluation, and \Visualization.
}
Each stage the machine executes enriches the data. The enriched data are summarized as data patterns and delivered by visualizations. 
We detail the labors of machines while we detail each stage in \autoref{sec:stages}.

\fussy

\subsection{Workflow Stages}

\label{sec:stages}

\sloppy
We outline six workflow stages of our model. The stages' composition draws inspiration from the KDD process model \cite{fayyad96aaai}, which describes how raw data passes through preprocessing, transformation, and data mining to yield final visualizations. Here, we substitute the transformation and data mining steps to \DR and \Evaluation stages to better match with visual analytics workflow using DR. 
We thus describe the stages in the sequence that are aligned with the KDD process. 
It is still important to note that the nature of interactive visual analytics causes the execution of these stages to be nonlinear: \analysts can interact with machines to update configurations, which can cycle the analysis back to the \machine-side stages (\autoref{sec:analyst}).

\fussy

\paragraph{\Preprocessingbf}
The preprocessing stage involves a variety of data manipulations that are executed before applying the DR techniques. The step receives raw \textit{high-dimensional data} as input and outputs the \textit{preprocessed data} based on the configurations specifying preprocessing operations to execute.
It can be largely divided into row operations and column operations. Data imputation, subsampling, or outlier removal are representative row operations. Regarding column operations, attribute selection that samples out or less weighs non-significant attributes is widely adopted. Attribute selection can also be conducted automatically, e.g., to reveal hidden patterns \cite{fujiwara23pacificvis} or maximize specific patterns like class separation \cite{wang18tvcg, sips09cgf}. Another typical column operation is to introduce new columns that represent the original structure's feature, e.g., by executing clustering techniques and adding the resulting clusters as labels to the datasets \cite{wenskovitch20iui, wenskovitch18tvcg}.

\paragraph{\DimRedbf or \DRbf}
This stage gets the \textit{preprocessed data} as input and executes DR algorithms, producing \textit{DR projections} as outputs. The stage is controlled by five configurations: DR techniques, technique-specific hyperparameters, initialization method, objective function (i.e., loss function), and distance function. The first two configurations should be necessarily set to execute the stage. The last three configurations are widely set for nonlinear DR techniques (e.g., UMAP, $t$-SNE, and Isomap \cite{tenenbaum00science}), which work by optimizing the 2D positions of data points to preserve the original distances between them.
The initialization method defines how the points will be positioned before starting optimization, and the objective function defines how the preservation of distances will be quantified. 
These configurations substantially impact the resulting DR projections. 
For example, Kobak et al. \cite{kobak21nb} show that initializing points using PCA before the optimization leads to more accurate DR results than using random initialization. Lee and Verleysen \cite{lee11pcs} show that the design of the distance function affects the projection accuracy.

\paragraph{\Evaluationbf}
This stage gets \textit{high-dimensional data}, \textit{preprocessed data}, and \textit{DR projections} as inputs and produces evaluation results as numerical \textit{scores}. These scores (1) explain how well the projections support the analytic tasks of investigating high-dimensional data or (2) the validation of hypotheses.
For example, when analysts want to conduct cluster analysis, it is important to evaluate the projections in advance using the evaluation metrics assessing the preservation of cluster structure \cite{sips09cgf, jeon21tvcg}.
This stage is affected by two configurations: metrics and metric hyperparameters. For example, Trustworthiness \& Continuity \cite{venna06nn}, which examine how well DR projections preserve the local neighborhood structure of original data, require a hyperparameter that designates the number of nearest neighbors to be considered.
\revise{
Note that these quality metrics differ from ``perceptual metrics'' (e.g., Scagnostics \cite{dang14pvis} or the class separation measure \cite{aupetit15pvis}), which focus on the perceived patterns of 2D projections without accounting for the original high-dimensional (HD) data. Please refer to Behrisch et al. \cite{behrisch18cgf} for detailed descriptions on these metrics.
}

\paragraph{\Visualizationbf}
\machines get all data cascaded from the previous stages (raw \textit{high-dimensional data}, \textit{preprocessed data}, \textit{projections}, and \textit{scores}) as input, generating a visualization or a set of visualizations delivering \textit{data patterns}. The main configuration that affects the patterns is the visualization methods for depicting DR projection. This incorporates the selection of not only the type and number of visual idioms (e.g., 2D or 3D scatterplots \cite{sedlmair13tvcg}) but also displays (e.g., 2D screen or VR headsets \cite{in24chi}) or the use of animations \cite{asimov85siam}. The way of encoding axes is another configuration that affects the knowledge generation process. For example, when the projection is made by selecting a few subspaces in \Preprocessing step, such information can be encoded in axes. Configuring how monochrome scatterplots can be augmented also substantially affects the resulting data patterns. Color-encoding the classes, for example, could cause analysts to overlook the separation between intrinsic clusters within the data \cite{sedlmair12cgf}. Also, how and how much projections suffer from distortions can be augmented by encoding the scores from the evaluation on the projections \cite{aupetit07neurocomp, lespinats07tnn, jeon21tvcg}, making analysts be more aware of distortions. Finally, auxiliary visualizations can be configured to make the original high-dimensional data more interpretable \cite{chatzimparmpas20tvcg, yan24pvis, kwon17tvcg}.



\paragraph{\Sensemakingbf}
\analysts make sense of data by investigating \textit{data patterns} delivered by the \Visualization. The procedure starts by perceiving the patterns, e.g., cluster \cite{jeon24tvcg2, abbas19cgf} or local neighborhood \cite{lespinats11cgf, aupetit05neurocomputing} patterns, aligned with the target task. The perception usually relies on the Gestalt principle of proximity and similarity \cite{palmer99gestalt}, where analysts perceive data points that have high proximity or similar color to be closely located in the original high-dimensional space.

\paragraph{\Interactionbf}
 \analysts can also interact with data, meaning they ``signal'' \machines to update configurations, thereby updating data patterns to align with their intentions. 
To do so, \analysts first evaluate whether the data patterns are suitable for validating hypotheses. If not suitable, they can interact with \machines to correct the visualizations. For instance, when analysts notice that the evaluation scores of DR projections indicate insufficient accuracy to properly support the target task, they can reconfigure the DR techniques or adjust hyperparameters to correct the projections. The \machines then again process the data based on the updated configuration, which leads to a new cycle of sensemaking. 




\subsubsection*{Model applications} In Appendix A, we verify the completeness and applicability of our workflow model by instantiating it on two research works, contributing interactive visualizations leveraging DR: \textit{AxiSketcher} \cite{kwon17tvcg} and \textit{CommentVis} \cite{yan24pvis}.

\section{Taxonomy}

\label{sec:taxonomy}

We derive four dimensions---\stages, \problems, \aims, and \solutions---that explain existing research papers that address the unreliability of visual analytics using DR. 
Our taxonomy builds upon the workflow model (\autoref{sec:workflow}): every branch of the taxonomy is incorporated as an element in the workflow model or can be explained using the elements. 
Our taxonomy provides a solid foundation for organizing and classifying relevant papers (\autoref{tab:taxonomy1}), thereby helping practitioners gain a clearer understanding of the field.

\subsection{\stages}

This dimension designates which workflow stage (\autoref{sec:stages}) causes unreliability. 
This dimension 
encompasses four \machine-side stages: \Preprocessing, \DR, \Evaluation, and \Visualization. 
Note that we do not include the \analyst-side stages (\Sensemaking and \Interaction) because unreliable visual analytics implies these stages to be problematic. 
In other words, all papers we identify face issues in the \analyst-side stages.
For example, if DR projections become inaccurate due to distortions, this leads to inaccuracies in both the sensemaking of the original structure and the interaction with the projection (e.g., the clusters brushed by \analysts may not represent actual clusters in the original space).
Please refer to \autoref{sec:stages} for detailed explanations of these stages.

\revise{Note that the stages where the unreliability is caused may not necessarily align with the stages where the \solutions are implemented. For example, accuracy issues in a dimensionality reduction (\DR) technique may be addressed through the development of novel \DR approaches \cite{jeon22vis, joia11tvcg} or by employing auxiliary visual encodings to demonstrate where and how inaccuracies occur.  }

\subsection{\problems}

\label{sec:problem}

This dimension details why visual analytics using DR become unreliable. This dimension consists of eight classes. 


\paragraph{Inaccurate}
A stage is considered inaccurate when errors occur during its execution, even though \analysts have set configurations that align with their intentions.
 In terms of \DR stage, this problem is widely represented by the term ``distortion'' \cite{aupetit07neurocomp, jeon21tvcg, lespinats07tnn, nonato19tvcg}, which is frequently characterized as an unavoidable \cite{sips09cgf, jeon24tvcg} error in projection arising from the intrinsic complexity of high-dimensional data.
  \revise{
 For example, Lespinats and Aupetit \cite{lespinats11cgf} address distortions of neighborhood structure in DR projections, and Martins et al. \cite{martins14cg} focus on the distortions that occur at the cluster level. 
 }
 Few papers state that \Evaluation can also be unreliable due to inappropriate design of DR evaluation metrics \cite{jeon24tvcg} or study design \cite{aupetit14beliv}.

\paragraph{Suboptimal}
Papers addressing suboptimality problems focus on situations where a particular configuration does not align with the target task. For example, Narayan et al. \cite{narayan21nature} and Jeon et al. \cite{jeon22vis} improve UMAP to better support the analytic tasks that examine global structure (e.g., cluster arrangement, pairwise distances between points). Similarly, Meng et al. \cite{meng24tvcg} and Hajderanj et al. \cite{hajderanj19icsie} claim that the original $t$-SNE does not well fit to classification task and thus improved it be supervised. Some papers also tackle the suboptimality in \Visualization stage. For instance, Bian et al. \cite{bian21tvcg} improve the conventional scatterplot to better support the exploration of subspace by changing the encoding of dots representing data points.

\paragraph{Incomplete}
If analysts cannot find a single configuration that matches their intention, the corresponding stage is incomplete. 
This problem mainly occurs during \Evaluation stage. 
Sips et al. \cite{sips09cgf} and Friedman et al. \cite{friedman87jasa} account for the problem that no existing DR evaluation metric examines class separability. Johansson and Johansson \cite{johansson09tvcg} propose weighting DR evaluation metrics to make the evaluation complete against arbitrary user tasks. The problem also occurs in \DR stage, 
where existing DR techniques cannot cover diverse analytic tasks \cite{friedman87jasa, lai18vlc}.


\paragraph{Unstable}
A stage is unstable if the output (e.g., projections in \DR stage) varies as configuration changes. 
For example, nonlinear DR techniques like $t$-SNE and UMAP can produce different projections when different hyperparameter settings or initialization methods are used \cite{fadel15neurocomp,kobak21nb}. 
The randomness in the optimization procedure can also make DR techniques unstable \cite{wattenberg2016tsnetuning, jung23vis}.
The papers thus make \DR stage more robust to such changes \cite{fadel15neurocomp} or recommend the best configuration \cite{kobak21nb}. The researchers also provide similar endeavors to remedy instability in DR evaluation metrics (\Evaluation stage) \cite{johannemann19arxiv, angelini22tvcg, lee10prl}.

\paragraph{Uninterpretable}
A stage is less interpretable if (1) the underlying mechanism of the stage or (2) the input data to the stage is not well explained to the \analysts. 
The former problem widely occurs in \DR and \Evaluation stages due to the intrinsic complexity of DR techniques \cite{choo10vast, chatzimparmpas20tvcg} or evaluation metrics \cite{lee09neurocomp}. 
The latter problem mainly stems from \Visualization stage. 
For example, scatterplots depicting DR projections can hardly inform the original attribute values to the analysts \cite{kwon17tvcg, faust19tvcg, dowling19tvcg}. The scatterplots also mostly represent data items using dots, which have problems depicting the original data format, e.g., text \cite{kandogan12vast}.

\paragraph{Unscalable against dimensionality}
The unscalability problem occurs when the complexity of final data patterns grows with increasing dimensionality, exceeding the available display pixels or human cognitive capabilities. For example, \DR stage can be unscalable for subspace analysis as the number of subspaces to explore increases exponentially against dimensionality \cite{friedman87jasa, leban06dmkd, turkay11tvcg, nam13tvcg}. Some papers claim that the unscalability originates from the \Visualization stage due to the fundamental limitation of scatterplots: each scatterplot can only depict a single projection at once \cite{self18tiis, jackle17vast}. 

\paragraph{Irreflective of domain knowledge}
If a stage hardly reflects the domain knowledge of \analysts, it means that resulting data patterns (external representation of data) do not align with the internal cognitive map of the analysts \cite{archambault13ijhci,orient12arch}. Kim et al. \cite{kim16tvcg} and Kwon et al. \cite{kwon17tvcg} address the problem that \Visualization stage becomes irreflective of domain knowledge as static scatterplots cannot receive user input. Xia et al. \cite{xia23tvcg} and Brown et al. \cite{brown12vast} claim that the problem originates from the \DR stage as DR techniques are executed independently with domain knowledge.

\paragraph{Uninformed}
\analysts are uniformed to a stage if they are unaware of (i.e., lack operational knowledge on) diverse techniques used in the stage. Every stage can thus be uninformed to analysts. Analysts are often uninformed about the other reliability problems, e.g., suboptimal \Evaluation \cite{aupetit14beliv, jeon24tvcg}. The difference between \textbf{uninformed} and \textbf{uninterpretable} is that the former can lead analysts to establish inappropriate configurations, while the latter incurs unreliability during \Sensemaking and \Interaction.

\subsection{\aims}

The \aim dimension is the ultimate objective that each paper seeks to accomplish. \revise{While the first class (Enhance reliability) aims to directly ``fix'' the problem, the latter two classes (Enhance awareness, Enhance approachability) purpose to empower \analysts to help them address the problem on their own.}

\paragraph{Enhance reliability}
The papers in this category aim to enhance reliability by providing newly designed techniques regardless of a stage or a problem \cite{jeon22arxiv, alcaide21tvcg}. Some papers do so by introducing the improved version of the existing techniques \cite{jeon22vis, narayan21nature, meng24tvcg}.


	\textsand{
	\begin{table*}[htbp]
		\centering
		\caption{\textit{The list of 133 papers we survey (rows) and their classification based on the proposed taxonomy (columns).} Papers within each group and type are arranged in descending order of their total citation as of September 2024, based on Google Scholar.}
  \Description{This table provides an overview of 133 surveyed papers (rows) and their classification based on a proposed taxonomy (columns). The table is divided into several sections: Stage: Papers are categorized based on whether they address preprocessing, dimensionality reduction, quantitative evaluation, or visualization. Problem: Issues such as inaccurate, suboptimal, incomplete, unstable, uninterpretable, unscalable, irreflective, and uninformed methods are marked. Aim: The aims include enhancing reliability, awareness, and approachability. Solution: Solutions are classified into improvements on DR, evaluation, visualization stage solutions, DR frameworks, visual analytics systems, literature reviews, human-centered experiments, and computational experiments. Each paper is marked with a checkmark in relevant columns, representing their contributions across these categories. The papers are sorted in descending order of citation count as of September 2024.}
		
		\label{tab:taxonomy1}
		\scalebox{0.77}{
\begin{tabular}{|c|c|rr|@{\hskip 2pt}|c?c?c?c|c?c?c?c?c?c?c?c|c?c?c|c?c?c?c?c?c?c?c|}
			\arrayrulecolor{black}\cline{5-8}\arrayrulecolor{black}\cline{9-16}\arrayrulecolor{black}\cline{17-19}\arrayrulecolor{black}\cline{20-27}
		\multicolumn{4}{c|}{} & \multicolumn{4}{|c|}{\textbf{Stage}} & \multicolumn{8}{|c|}{\textbf{Problem}} & \multicolumn{3}{|c|}{\textbf{Aim}} & \multicolumn{8}{|c|}{\textbf{Solution}} \\
		\arrayrulecolor{black}\cline{5-8}\arrayrulecolor{black}\cline{9-16}\arrayrulecolor{black}\cline{17-19}\arrayrulecolor{black}\cline{20-27}
		\multicolumn{4}{c|}{} & \rotatebox{90}{Preprocessing} & \rotatebox{90}{Dimensionality Reduction} & \rotatebox{90}{Quantitative Evaluation} & \rotatebox{90}{Visualization} & \rotatebox{90}{inaccuate} & \rotatebox{90}{suboptimal} & \rotatebox{90}{incomplete} & \rotatebox{90}{unstable} & \rotatebox{90}{uninterpretable} & \rotatebox{90}{unscalable} & \rotatebox{90}{irreflective} & \rotatebox{90}{uninformed} & \rotatebox{90}{Enhance reliability} & \rotatebox{90}{Enhance awareness} & \rotatebox{90}{Enhance Approchability} & \rotatebox{90}{Improvements on DR} & \rotatebox{90}{Improvements of Evaluation} & \rotatebox{90}{Visualization Stage Solutions} & \rotatebox{90}{DR framework} & \rotatebox{90}{Visual Analytics System} & \rotatebox{90}{Literature Review} & \rotatebox{90}{Human-Centered Experiment} & \rotatebox{90}{Computaional Experiment} \\
		\arrayrulecolor{black}\hline
		\cellcolor{applepinknormal!20} & 1 & Roweis and Saul \cite{roweis00science} & 2000 &  & \cellcolor{Goldenrod} \color{white}{\checkmark} &  &  &  & \cellcolor{Goldenrod} \color{white}{\checkmark} &  &  &  &  &  &  & \cellcolor{Goldenrod} \color{white}{\checkmark} &  &  & \cellcolor{Goldenrod} \color{white}{\checkmark} &  &  &  &  &  &  &  \\
\hline

		\end{tabular}}
	\end{table*}
	}



	\textsand{
	\begin{table*}[htbp]
		\centering
		\label{tab:taxonomy2}
		\scalebox{0.77}{
		\begin{tabular}{|c|c|rr|@{\hskip 2pt}|c?c?c?c|c?c?c?c?c?c?c?c|c?c?c|c?c?c?c?c?c?c?c|}
			\arrayrulecolor{black}\cline{5-8}\arrayrulecolor{black}\cline{9-16}\arrayrulecolor{black}\cline{17-19}\arrayrulecolor{black}\cline{20-27}
		\multicolumn{4}{c|}{} & \multicolumn{4}{|c|}{\textbf{Stage}} & \multicolumn{8}{|c|}{\textbf{Problem}} & \multicolumn{3}{|c|}{\textbf{Aim}} & \multicolumn{8}{|c|}{\textbf{Solution}} \\
		\arrayrulecolor{black}\cline{5-8}\arrayrulecolor{black}\cline{9-16}\arrayrulecolor{black}\cline{17-19}\arrayrulecolor{black}\cline{20-27}
		\multicolumn{4}{c|}{} & \rotatebox{90}{Preprocessing} & \rotatebox{90}{Dimensionality Reduction} & \rotatebox{90}{Quantitative Evaluation} & \rotatebox{90}{Visualization} & \rotatebox{90}{inaccuate} & \rotatebox{90}{suboptimal} & \rotatebox{90}{incomplete} & \rotatebox{90}{unstable} & \rotatebox{90}{uninterpretable} & \rotatebox{90}{unscalable} & \rotatebox{90}{irreflective} & \rotatebox{90}{uninformed} & \rotatebox{90}{Enhance reliability} & \rotatebox{90}{Enhance awareness} & \rotatebox{90}{Enhance Approchability} & \rotatebox{90}{Improvements on DR} & \rotatebox{90}{Improvements of Evaluation} & \rotatebox{90}{Visualization Stage Solutions} & \rotatebox{90}{DR framework} & \rotatebox{90}{Visual Analytics System} & \rotatebox{90}{Literature Review} & \rotatebox{90}{Human-Centered Experiment} & \rotatebox{90}{Computaional Experiment} \\
		\arrayrulecolor{black}\hline
		\cellcolor{applegreennormal!20} & 66 & Van der maaten and Postma \cite{vandermaaten09jmlr} & 2009 &  & \cellcolor{Goldenrod} \color{white}{\checkmark} &  &  &  &  &  &  &  &  &  & \cellcolor{Goldenrod} \color{white}{\checkmark} &  & \cellcolor{Goldenrod} \color{white}{\checkmark} &  &  &  &  &  &  &  &  & \cellcolor{Goldenrod} \color{white}{\checkmark} \\
\hline

		\end{tabular}}
	\end{table*}
	}



\begin{figure*}
    \centering
    \includegraphics[width=\linewidth]{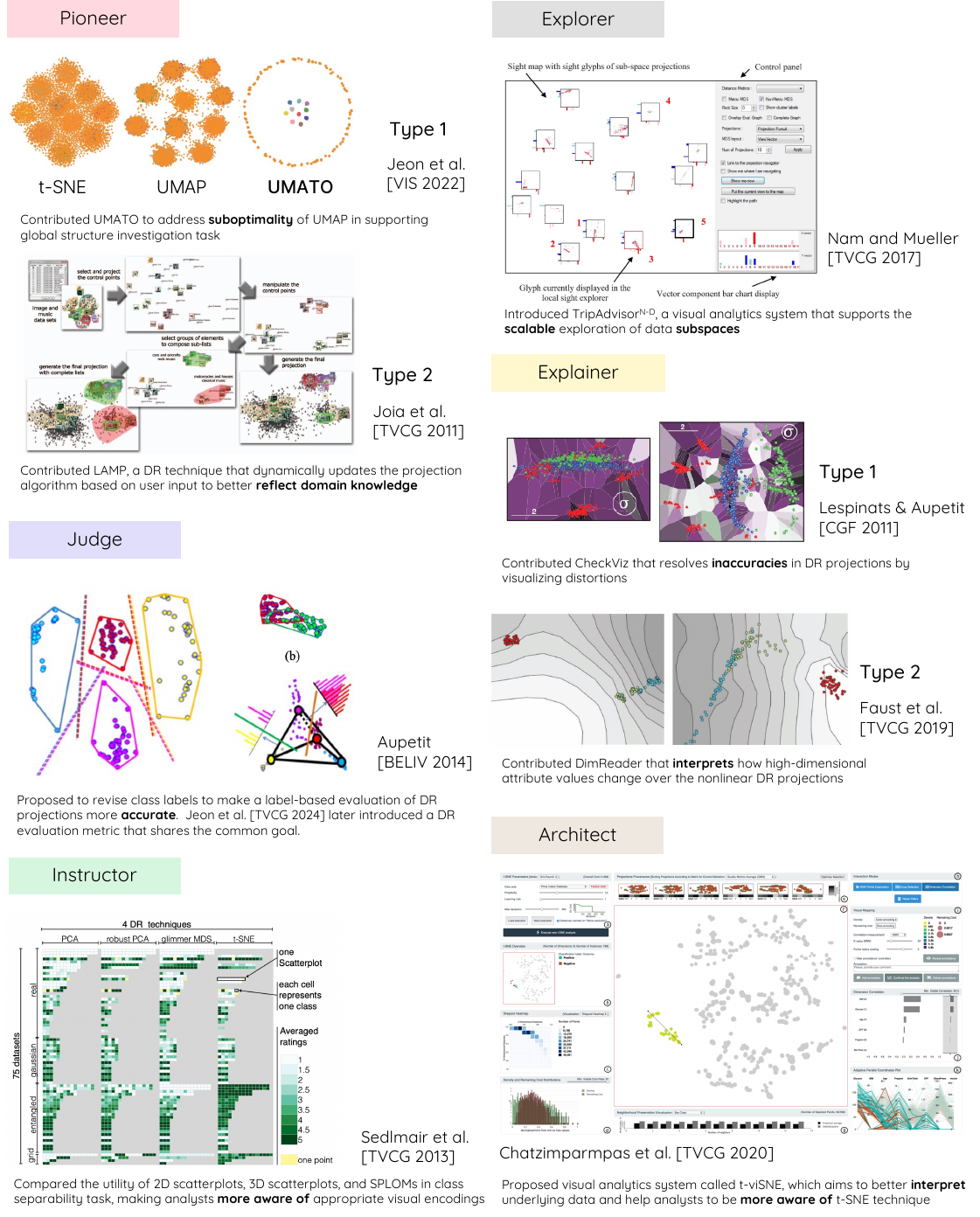}
    \caption{\textit{Example papers in each cluster we identify by conducting meta-analysis} (\autoref{sec:metaanalysis}). The problems (\autoref{sec:problem}) that each system resolves are highlighted in bold.  Reference papers: \Pioneer \cite{jeon22vis, joia11tvcg},  \Judge \cite{aupetit14beliv}, \Instructor \cite{sedlmair13tvcg}, \Explorer \cite{nam13tvcg},  \Explainer \cite{faust19tvcg, lespinats11cgf}, and \Architect \cite{chatzimparmpas20tvcg}.}
    \label{fig:example_cluster}
    \Description{This figure provides example papers grouped into six clusters identified through a meta-analysis. Each cluster highlights specific problems the systems address, with corresponding solutions and visuals: Pioneer: Type 1: Jeon et al. [VIS 2022] introduced UMATO, addressing the suboptimality of UMAP in supporting global structure investigation tasks. Type 2: Joia et al. [TVCG 2011] contributed LAMP, a dynamic DR technique that updates projections based on user input to better reflect domain knowledge. Explorer: Nam and Mueller [TVCG 2017] proposed TripAdvisor, a scalable system for exploring data subspaces. Explainer: Type 1: Lespinats and Aupetit [CGF 2011] introduced CheckViz to resolve inaccuracies in DR projections through distortion visualization. Type 2: Faust et al. [TVCG 2019] developed DimReader, which interprets how high-dimensional attribute values change across nonlinear DR projections. Judge: Aupetit [BELIV 2014] proposed revisions to class labels for more accurate DR projection evaluations. Jeon et al. [TVCG 2024] later introduced a related evaluation metric. Instructor: Sedlmair et al. [TVCG 2013] compared scatterplots and SPLOMs, making analysts more aware of appropriate visual encodings. Architect: Chatzimparmpas et al. [TVCG 2020] proposed the v-tSNE visual analytics system to better interpret underlying data and make analysts more aware of t-SNE technique applications.
}
\end{figure*}


\paragraph{Enhance awareness}
The papers inform \analysts to be more aware of the possible configurations, addressing the problem that the stages are \textbf{uninterpretable} \cite{jackle17vast, kwon17tvcg} or \textbf{uninformed} \cite{nonato19tvcg, xia22tvcg}. This incorporates the efforts of making \analysts more aware of all kinds of reliability problems, e.g., \textbf{inaccuracy} \cite{aupetit07neurocomp} and \textbf{suboptimality} \cite{cutura20avi}.

\paragraph{Enhance approachability}
These papers aim to make \analysts easier to \textbf{interpret} techniques \cite{chatzimparmpas20tvcg, cavallo18chi}, handle the \textbf{instability} of the techniques \cite{molchanov20pacificvis}, and more easily inject their \textbf{domain knowledge} \cite{kwon17tvcg, kim16tvcg}. 

\subsection{\solutions}

\label{sec:solution}

This dimension explains the contribution of the papers that materializes their \aim. The dimension is classified into eight classes. 
\revise{Note that the first three categories (Improvements in \Preprocessing, \Evaluation, and \Visualization) differ from the categories in the \stages; the former indicates where reliability is addressed, while the latter identifies where the reliability issues originate.
}
There is no class entitled ``Improvements in \Preprocessing stage'' as we find no papers that provide solutions that improve \Preprocessing stage.

\paragraph{Improvements in \DR stage}
Papers in this category provide solutions for improving the DR stage, thus addressing the problems in the same stage. This includes the introduction of new DR techniques \cite{fadel15neurocomp, spathis18eccv}, improvement of DR techniques \cite{jeon22vis, meng24tvcg, vu21ijcnn}, or the improvement of a certain subpart of DR techniques, e.g., initialization \cite{kobak21nb} or distance functions of nonlinear DR techniques \cite{brown12vast}. 
\revise{Jeon et al. \cite{jeon22vis}, for instance, addresses the \textbf{inaccuracy} and \textbf{suboptimality} of UMAP in representing global structure by altering the optimization process.}

\paragraph{Improvements in \Evaluation stage}
As with the previous category, these papers here provide solutions related to the Quantitative Evaluation stage, aiming to address problems in the Evaluation and previous stages. For instance, the papers introduce new DR evaluation metrics both to address \textbf{inaccuracy} of \DR or \Preprocessing stage \cite{motta15neurocomp} or to resolve \textbf{incompleteness} or \textbf{instability} of the Evaluation stage \cite{jeon21tvcg, lee09neurocomp}. 
Another case is a library offering DR evaluation metrics, which improves their \textbf{approachability} \cite{jeon23vis}.

\paragraph{Improvements in \Visualization stage}
These papers improve the Visualization stage, addressing problems in not only the Visualization but also previous stages. One notable types of paper in this category depict where and how much the DR projections are distorted by overlying such info over projections \cite{lespinats11cgf, jeon21tvcg} or dynamically relocating points to ``correct'' the distortions \cite{jeon22arxiv, heulot13vamp, stahnke16tvcg}, dealing with \textbf{inaccuracy} in \DR stage.

\paragraph{DR framework}
These papers contribute a framework that automatically specifies configurations across multiple stages. For example, Choo et al. \cite{choo09vast} propose a framework that establishes a configuration that mixes two DR techniques, aiming to address each technique's \textbf{suboptimality}. Fujiwara et al. \cite{fujiwara23pacificvis} aim to find a set of configurations in \Preprocessing and \DR stage that maximizes the diversity of the patterns found in subspace analysis.

\paragraph{Visual analytics system}
The papers in this category contribute a visual analytics system, 
These works often address \textbf{interpretability} issues of \DR or \Visualization stage to \textbf{enhance awareness} of DR techniques \cite{chatzimparmpas20tvcg, cutura18esann, jeong09cgf} and resulting projections \cite{xia17tvcg}.
We classify the papers to \textbf{improvements in \Visualization stage} category if they are more focused on proposing novel visualization design and classify them to this category if their main contribution is a dashboard system consisting of linked visualizations.

\paragraph{Literature review}
The researchers contribute literature review on DR techniques (\DR stage) \cite{nonato19tvcg, engel12irtg} and evaluation metrics (\Evaluation stage) \cite{jeon23vis, nonato19tvcg}, mainly aiming to \textbf{enhance awareness} of \analysts by informing their characteristics. 
These papers serve as "guidelines" for \analysts in setting up appropriate configurations.
We can interpret the contribution of these papers as the authors' effort to share their operational knowledge.

\paragraph{Human-centered experiment}
This type of research work is similar to \textbf{literature reviews}, but provides evidence about their guidelines by conducting user studies \cite{etemadpour15tvcg, xia22tvcg}. 
For example, they determine the usability of DR techniques based on human task performance on their projections.
Their guidelines are thus easy to apply in practice: configurations are directly mapped to suitable tasks. 
These papers can be interpreted as the authors performing \Evaluation and \Sensemaking to gain and share operational knowledge.

\paragraph{Computational experiment}
The papers in this category serve the same role as those in the \textbf{human-centered experiment} category, except that the evaluation of DR techniques is conducted computationally using DR evaluation metrics. Their guidelines are thus less applicable compared to the ones from human-centered experiment. For example, even if we use DR evaluation metrics that examine the preservation of cluster structure (e.g., Steadiness \& Cohesiveness), we are not sure about whether metric scores directly align with human performance in conducting cluster analysis.

\section{Meta-Analysis on the Taxonomy}

\label{sec:metaanalysis}

We systematically investigate how the classified papers (\autoref{tab:taxonomy1}) can be categorized at a higher level.
Our analysis results help practitioners quickly overview the papers we identify, thereby reducing their cognitive load to understand the research landscape and make their visual analytics more reliable.
The analysis findings also serve as a foundation for identifying ongoing research challenges (\autoref{sec:challanges}) and imbalances in the research landscape (\autoref{sec:implication}, \ref{sec:mitigation}).

\subsection{Objectives and Design}

We group papers that are similarly classified according to our taxonomy (\autoref{sec:taxonomy}).
We follow the data-driven approach of clustering papers proposed by the survey conducted by Shin et al. \cite{shin23tvcg}.

\paragraph{Cluster identification}
We first convert the categorization of papers into a data table. The rows of the data table correspond to the papers, and the columns correspond to each category in our taxonomy (\autoref{tab:taxonomy1}). Each paper is thus represented as a binary vector, where each element is 1 if the paper belongs to the corresponding category and 0 if it does not.

We then apply three different clustering algorithms (hierarchical clustering \cite{murtagh12wires}, DBSCAN \cite{kriegel11dmkd}, and HDBSCAN \cite{campello13kdd}) while using Jaccard distance to measure the dissimilarity between data points. 
To find the optimal clustering results, we test various hyperparameter settings with Bayesian optimization \cite{snoek12nips} while using Silhouette coefficient \cite{rousseuw87silhouette} as a target variable. 
The total number of iterations for Bayesian optimization is 100. We select the clustering results with the best Silhouette coefficient score. 
We use \texttt{scikit-learn} implementation (\texttt{AgglomerativeClustering}) \cite{pedregosa11jmlr} for clustering algorithms and Nogueira's implementation \cite{nogueira14github} for the Bayesian optimization. 

\paragraph{Manual validation}
As clustering algorithms do not always output optimal clustering results \cite{bendavid08neurips}, two authors manually screen the papers to validate the individual compactness and mutual separability of the clusters \cite{bendavid08neurips}. The authors revise the grouping by merging clusters that have papers with similar characteristics (See \autoref{sec:clusterresults} for results).

\subsection{Identified Clusters}

\label{sec:clusterresults}

We find that hierarchical clustering yields the best result, consisting of seven clusters with a Silhouette score of 0.292.
After manual validation, the authors merge two clusters with papers with similar purposes, resulting in six clusters: 
\Pioneer, \Judge, \Instructor, \Explorer, \Explainer and \Architect.
The following are detailed explanations of these clusters, characterized by their role in visual analytics using DR. Please refer to \autoref{tab:taxonomy1} to the list of papers within each cluster (linked by color). Also refer to \autoref{fig:example_cluster} for the example figures of the papers within each cluster we discuss.

\paragraph{\Pioneerbf}
Papers in this cluster ``pioneer'' reliable visual analytics by contributing methodologies to produce better DR projections. 
One branch of papers within this cluster proposes new DR techniques \cite{matute20cgf, joia11tvcg} or improves the existing DR techniques \cite{geng05tsmc, jeon22vis, meng24tvcg} to address \textbf{incompleteness}, \textbf{inaccuracy}, and \textbf{suboptimality} in \DR stage (\textbf{type 1}; row 1--33 in \autoref{tab:taxonomy1}). 
For instance, Jeon et al. \cite{jeon22vis} address the limitations of UMAP in supporting global structure investigation tasks. 
These papers contribute static DR techniques, which means that they are not updated or affected by \Interaction.
In contrast, the remaining papers propose DR techniques that interactively update the inner logic based on user input (\textbf{type 2}; row 34--48). For example, Joia et al. \cite{joia11tvcg} propose to dynamically update the projection algorithm by reflecting the user interaction that updates the 2D positions of points.

\paragraph{\Judgebf}
The papers in this cluster address problems regarding \Evaluation stage by proposing new DR evaluation metrics or strategies, supporting \analysts to reliably judge the quality of DR projections.
For example, Aupetit \cite{aupetit14beliv} reveals \textbf{inaccuracy} in evaluating cluster structure of projections by using class labels as ground truth clusters. The paper not only makes aware of the issue but also remedies it by proposing to revise the class labels to better reflect cluster structure.
Jeon et al. \cite{jeon24tvcg} later provide a more tangible solution to this issue by introducing new DR evaluation metrics.

\fussy

\paragraph{\Instructorbf}
The papers in this cluster teach \analysts (i.e., make them \textbf{aware} of) how to establish effective configurations, mostly for \DR stage.
These papers mostly contribute \textbf{literature reviews}, \textbf{human-centered experiment}, and \textbf{computational experiment}. One exception is Cutura et al. \cite{cutura18esann}, where the authors contribute to a \textbf{visual analytics} system that explains and compares different DR techniques. A few papers in this cluster also guide setting effective configurations in \Visualization stage. For example, Sedlmair et al. \cite{sedlmair13tvcg} conduct an experiment comparing different visualization types (2D scatterplot, 3D scatterplot, and SPLOM) in supporting the class separability tasks using DR projections.

\paragraph{\Explorerbf}
These papers aim to make the exploration of subspaces more \textbf{approachable}. These papers address the \textbf{scalability} problem, which arises from the increasing number of subspaces that need to be investigated as dimensionality increases. For example, Asimov \cite{asimov85siam} propose to animate scatterplots to allow analysts to ``tour'' the possible set of subspaces, and Nam and Mueller \cite{nam13tvcg} elaborate this approach based on the metaphor of tourism in our real life. 

\paragraph{\Explainerbf}
The papers here mostly address \textbf{inaccuracy} or \textbf{uninterpretability} of \DR or \Visualization stages by augmenting scatterplots. In terms of accuracy, this is typically done by overlaying how and where projections are distorted \cite{aupetit07neurocomp, martins15svcg, lespinats11cgf} or resolving distortions by moving them \cite{heulot13vamp, jeon22arxiv} (\textbf{type 1}; rows 94--102). Similarly, interpretability is addressed by overlying high-dimensional attribute values to the scatterplots \cite{faust19tvcg} (\textbf{type 2}; rows 103--111).

\paragraph{\Architectbf}
These papers architect \textbf{visual analytics systems} that help people better understand DR techniques or underlying data. For example, Jeong et al. \cite{jeong09cgf} and Chatzimparmpas et al. \cite{chatzimparmpas20tvcg} contribute visual analytics systems that enhance \analysts{}'s awareness of the inner logic of PCA and $t$-SNE, respectively. On the other hand, Xia et al. \cite{xia17tvcg} and Liu et al. \cite{liu14cgf} contribute visual analytics systems that help users conduct more reliable analysis despite DR techniques' inherent \textbf{incompleteness} and \textbf{instability}.

\subsection{Implication: Imbalance in the Research Landscape}

\label{sec:implication}

Our meta-analysis reveals the imbalance of research topics in the field. While we identify 48 \Pioneer-type papers (36.1\%), we find a relatively small number of papers for other types of papers. For example, there exists only 17 (12.8\%), and 22 (16.5\%) of \Judge-type papers and \Instructor-type papers. The situation is worse for \Explorer, as it only incorporates six papers. 
The finding suggests that the research community invests substantial effort in developing new DR techniques but comparatively less effort in evaluating them and making them interpretable. 
From the perspective of our taxonomy, this result suggests that while we have thoroughly studied the improvement of low-level performance in DR techniques (e.g., \textbf{inaccuracy}, \textbf{suboptimality}, \textbf{incompleteness}), the challenges associated with their high-level usage remain underexplored (e.g., \textbf{uninterpretability} and the \textbf{lack of reflection to domain knowledge}).

Such imbalance can potentially harm the reliability of visual analytics. For example, the absence of \Instructor-type papers can result in \analysts \textbf{being unaware of} newly proposed DR techniques, making it challenging to determine optimal configurations. 
We discuss tangible solutions to remedy such imbalance in \autoref{sec:challanges} and \autoref{sec:mitigation}.

\section{Research Challenges}

\label{sec:challanges}

After finishing the literature review and analysis, we again review the papers we identify and conduct a post-review discussion to reveal open challenges in the field. 
Six authors iteratively participate in the discussions.
As a result, we find three crucial challenges in the field. 

In this section, we first outline the challenges and suggest which direction the visualization research community should pursue to address them (\autoref{sec:challresearch}). 
We then detail expert interviews that verify the significance, urgency, and completeness of our challenges (\autoref{sec:challeval}).

\subsection{Challenges}

\label{sec:challresearch}

\sloppy
\paragraph{Challenge 1: We lack human-centered evaluations}
Human-centered evaluations provide applicable guidelines to \analysts (\autoref{sec:solution}). 
However, we have identified a lack of human-centered evaluations in the field.
Among 22 \Instructor-type papers, which is already small (\autoref{sec:implication}), only five papers contribute to human-centered experiments.

\fussy

We moreover identify that we lack recent human-centered evaluations. For \DR stage, while we have three papers contributing human-centered experiments to examine the utility of DR techniques in supporting analytics tasks, they either do not cover certain modern techniques (e.g., $t$-SNE or UMAP) \cite{etemadpour15tvcg} or limit their focus to specific analysis tasks (cluster analysis) \cite{xia22tvcg}.
The situation is similar for other stages (\Preprocessing, \Evaluation, and \Visualization); we need new experiments covering recently proposed techniques that can keep our guidelines up-to-date. 
For \Preprocessing stage, there is no human-centered evaluation at all. Addressing this problem is crucial as preprocessing, like normalizing data attribute values, significantly affects the resulting data patterns \cite{fujiwara23pacificvis}.
This finding highlights a need for a comprehensive human-centered experiment on modern popular techniques to offer more valuable insights to \analysts.

\noindent
\textit{Call for the research community.}
We suggest visual analytics researchers to conduct more human-centered experiments across all stages of our workflow model. 
We first highlight the importance of benchmarking recent DR techniques as they affect data patterns the most. 
We also want to prioritize the importance of conducting human-centered experiments on DR evaluation metrics, examining whether their scores align well with human task performance \cite{tatu10avi}. By knowing such alignment, we can approximate the potential human performance on different DR projections based on \textbf{computational experiments}, which is much more cost-efficient. This will thus enable practitioners to identify the most appropriate configurations for \DR stage that match the target task just by evaluating DR projections using the metrics. The endeavor will also contribute to mitigating the scarcity of \Instructor- and \Judge-type papers (\autoref{sec:implication}) by prompting practitioners to pay more attention to the importance of evaluations. 

We also recommend the community to incentivize replication studies. As mentioned in Quadri et al. \cite{quadri19arxiv}, visualization and HCI communities lack replication studies, which prohibits knowledge hardly maintained up-to-date or re-validated. Encouraging researchers to conduct more replication studies will strengthen the robustness of knowledge not only related to DR but also the ones relevant to other areas in visual analytics and HCI.
We thus also resonate with existing efforts on incentivizing replication studies\footnote{\url{https://ieeevis.org/year/2024/blog/vis-2024-OPC-blog-replication}}.

Finally, we suggest researchers build computational models that simulate human perception or cognition \cite{jeon24tvcg2, abbas19cgf, pandey16chi}.
These models cannot fully replace but still can approximate human-centered evaluations in a scalable and cost-effective manner. Moreover, such models can be integrated into the loss function of DR techniques, thus contributing to not only finding effective DR techniques but optimizing their projections \cite{jeon24tvcg2}.

\paragraph{Challenge 2: We require considerations of visual representations beyond 2D scatterplots}
While reviewing the list of identified papers, we notice that the visual representations used to depict DR projections are excessively bound to 2D scatterplots. 
2D scatterplots are certainly effective at showing spatial proximity between data points and also have high \textbf{approchability}, i.e., do not require high visualization literacy to understand. 
However, we also notify that many reliability problems originate from the use of 2D scatterplots in the \Visualization{} stage. As described in \autoref{sec:problem}, the visualization can be unreliable as 2D scatterplots are \textbf{suboptimal} or \textbf{less interpretable} \cite{bian21tvcg, cutura20avi}. 
We also find that the 2D scatterplots are \textbf{unscalable against dimensionality} \cite{asimov85siam, nam13tvcg}.

Moreover, our survey implies that scatterplots may not be the best visual representations to make visual analytics more explainable. In \Explainer cluster, \textbf{accuracy} and \textbf{interpretability} problems in DR techniques are addressed by augmenting 2D scatterplots (See \autoref{fig:example_cluster}). However, interpreting such augmented plots requires additional efforts. This barrier makes it difficult for these advanced visualization techniques to be applied in everyday data analysis.

\noindent
\textit{Call for the research community.}
We thus suggest investigating the utility of alternative visualization representations. 
One possibility is to investigate the utility of 3D scatterplots.
As 3D space has a higher degree of freedom compared to 2D space, 3D scatterplots can reduce unreliability due to the distortion of DR. 
Previously, Sedlmair et al. \cite{sedlmair13tvcg} verify that 3D scatterplots have no clear benefit compared to 2D scatterplots in representing cluster separability of DR projections in 2D displays, but other studies have shown different results with immersive analytics technologies: although not yet focused on DR projections and visual analytics, Yang et al. \cite{yang23tvcg} show that visual cues in immersive environments, especially motion cues, positively affect cluster perception of 3D scatterplots. 
Pursuing these efforts is also crucial as the effectiveness of immersive environments in data analysis for experts has just started to flourish \cite{in24chi}. 

We also recommend researchers investigate non-Euclidean layout space \cite{starnoneucl24}. Chen et al. \cite{chen21chi, chen22chi} show that mapping graph (i.e., network) data in a torus or sphere instead of a 2D plane significantly improves the accuracy in identifying clusters. Since graph layout can be interpreted as ``projecting'' data into a low-dimensional space \cite{kruiger17cgf}, these layout spaces can also be explored for DR. The machine learning community has made contributions in this area \cite{chami21icml}, but the utility of non-Euclidean spaces for DR has not yet been validated through user studies.

Finally, we encourage researchers to investigate the utility of visual idioms beyond scatterplots, such as parallel coordinates. To the best of our knowledge, this area has been largely unexplored and may hold significant potential for further study.

\paragraph{Challenge 3: We lack libraries}
In our survey, we find only one paper that introduces a library \cite{jeon23vis}, which implements a collection of DR evaluation metrics (\Evaluation stage). This is a surprisingly small number, particularly regarding the visualization community's acknowledgment of the success of libraries (e.g., D3 \cite{bostock11tvcg}, Vega-Lite \cite{satyanarayan17tvcg}, and ggdist \cite{kay24tvcg}) and their substantial contribution to the progress of research. The finding verifies that we lack libraries that serve diverse methodologies in a unified interface.

Outside of our survey scope, we have many libraries from the machine learning field for  \Preprocessing and \DR stages. For example, \texttt{scikit-learn} \cite{pedregosa11jmlr} provides diverse DR techniques\footnote{\url{https://scikit-learn.org/stable/modules/manifold.html}} and preprocessing functions\footnote{\url{https://scikit-learn.org/stable/modules/preprocessing.html}} using the Python interface. From a visualization perspective, Druid$_\text{JS}$ \cite{cutura20vis} supports four DR techniques in JavaScript, aligning with the trend of web-based development in the field.
However, these libraries only support widely used DR techniques (e.g., $t$-SNE, UMAP, Isomap), lacking support for diverse techniques contributed by the \Pioneer-type papers (e.g., LAMP \cite{joia11tvcg}, LoCH \cite{fadel15neurocomp}, UMATO \cite{jeon22vis}, NeRV \cite{venna10jmlr}). 
Still, to the best of our knowledge, no library exists to support the \Visualization stage.

More libraries will also help address the shortage of \Instructor-type papers (\autoref{sec:implication}), particularly those contributing to \textbf{human-centered} and \textbf{computational experiments}. For instance, a library that comprehensively encompasses DR techniques would greatly simplify the execution of controlled experiments comparing these techniques.

Therefore, although numerous methodologies to make visual analytics using DR reliable exist, these contributions can hardly be leveraged for practice. 
Few papers open-source their research artifacts \cite{fujiwara23pacificvis, jeon21tvcg, joia11tvcg, espadoto21tvcg, meng24tvcg}, but it is tedious to install and test these codes individually.
\analysts thereby may find it challenging to apply these techniques in practice. They might identify a suitable configuration that aligns with their intentions, perhaps by reading this paper, but will struggle to access and execute the corresponding techniques. To address this problem, we need more libraries that serve various techniques with a unified interface.

\noindent
\textit{Call for the research community.}
We advocate visual analytics researchers to build and deploy more libraries that materialize the methodologies contributed by the papers we survey. 
We especially recommend to build libraries that support \Visualization stages. This will help reduce the shortage of \Explorer, \Explainer, and \Architect-type papers by allowing researchers to easily test their new contributions with existing research artifacts.

However, similar to replication studies (see Challenge 1), implementing existing techniques is unlikely to be seen as novel by the community at this time. 
We thus incentivize the development of libraries of ``existing techniques'' as valid research contributions. This will significantly save the time of researchers and analysts, allowing them to dedicate more time and effort to do something truly ``novel.'' 
Here, we want to also emphasize the importance of avoiding over-reliance on libraries, as it can lead to unreliable research caused by coding errors. To avoid such a problem, not only deploying libraries but maintaining them should also be incentivized.

We additionally suggest making these libraries more unified. 
As seen in our workflow (\autoref{sec:workflow}), \analysts should frequently move back and forth between the stages. 
However, existing libraries support each stage individually, requiring analysts to integrate multiple libraries manually.
Providing libraries that allow analysts to establish and manage configurations of every stage simultaneously will substantially enhance the efficiency of data analysis.

\begin{table*}[t]
    \centering
    
    \caption{\textit{Demographic information of the expert interview participants.} These experts confirm the significance, urgency, and completeness of the research challenges we identify.
    \revise{``Exp.'', ``Pub.'' and ``Venue'' stands for the years of research experience, the number of archived publications related to or leverage dimensionality reduction published before Nov 2024 (according to Google Scholar), and the venues in which these papers are published, respectively. 
    The main research domains refer to the representative subareas of these DR-related papers.
    Also note that (A) and (I) stand for academia and industry, respectively.}}
    \Description{
    This table presents the demographic and professional details of eight expert participants (P1–P8) from academia and industry, highlighting their diverse roles, research expertise, and publication records. The participants include Assistant Professors, Senior Researchers, Engineers, and a Senior Lecturer, with ages ranging from 29 to 49 and research experience spanning 5 to 20 years. Their main research domains cover dimensionality reduction (DR) acceleration, evaluation, interpretation, interaction, clustering, explainable AI (XAI), bioinformatics, and machine learning. Publication counts vary from 3 to 20, with venues including TVCG, VIS, EuroVis, NeurIPS, CHI, and Distill. 
    }
     \scalebox{0.89}{
    \begin{tabular}{r|lllllll}
    \toprule
          &  \textbf{Occupation} &  \textbf{Gender} & \textbf{Age} & \textbf{Exp.} & \revise{\textbf{Pub.}} & \revise{\textbf{Main research domains}} & \revise{\textbf{Venue}}\\
    \midrule 
      P1  & Assistant Professor (Prof.) \revise{(A)} & Male &  32  & 10 & \revise{7} & \revise{DR acceleration and evaluation} & \revise{TVCG, VIS, EuroVis} \\   
      P2  & Senior Research Scientist \revise{(A)}   & Male & 49 & 20 & \revise{20} & \revise{DR evaluation and interaction} & \revise{TVCG, CGF, NeurIPS, Neurocomp., etc.} \\
      P3  & Assistant Prof. \revise{(A)}& Female & 31  & 7 & \revise{3} & \revise{Bioinformatics, XAI} & \revise{TVCG, CHI} \\ 
      P4  & Assistant Prof. \revise{(A)} & Male  & 29 & 7 & \revise{8} & \revise{DR Interpretation, XAI}  & \revise{TVCG, CGF, EuroVis, IVIS, PacificVis}\\ 
      P5  & Research Scientist, Adjunct Prof. \revise{(A)} & Male & 36 & 14 & \revise{6} & \revise{DR interaction, Clustering}  & \revise{TVCG, IUI, VDS, TiiS}\\ 
      \revise{P6} & \revise{Research Engineer (I)} & \revise{Male} & \revise{30}&  \revise{5} & \revise{11} & \revise{DR algorithm and interpretation} & \revise{TVCG, C\&G, IVIS, JoV, etc.}\\ %
      \revise{P7} & \revise{Software Engineer (I)} & \revise{Female} & \revise{30}& \revise{7} & \revise{4} & \revise{Machine learning, XAI} & \revise{NeurIPS, Distill} \\ 
      \revise{P8} & \revise{Assistant Prof. (A)} & \revise{Male} & \revise{40}& \revise{15}& \revise{17} & \revise{DR evaluation and interpretation} & \revise{TVCG, CGF, EuroVis, EuroVA, etc.}\\ 
      \bottomrule
    \end{tabular}
    }
    \label{tab:demo_expert}
\end{table*}

\begin{table*}[t]
    \centering
    
    \caption{\textit{The list of questions we asked experts to evaluate our challenges.} Significance and urgency are asked for each question, while completeness is assessed across all three challenges collectively. }
    \Description{
    This table outlines the questions asked to experts in order to evaluate the identified research challenges based on three criteria: Significance, Urgency, and Completeness. The table consists of three sections:  Significance: Q1: The challenge addresses an important problem within its field. Q2: Addressing this challenge has the potential to significantly advance current knowledge of our research field. Q3: Addressing this challenge could lead to significant improvements in practical applications. Urgency: Q4: The challenge is urgent in the context of current trends in research. Q5: Addressing this challenge will prevent negative consequences that can potentially happen in practice. Completeness: Q6: The list is complete; they cover important challenges that hinder making visual analytics more reliable. This table describes how significance and urgency were assessed for each challenge individually, while completeness was evaluated for all challenges collectively.
    }
    \scalebox{1.00}{
    \begin{tabular}{l|ll}
    \toprule
        \textbf{Criteria} & \textbf{No. }& \textbf{Questions}\\
    \midrule
        \multirow{3}{*}{Significance} &  Q1 & \textit{The challenge addresses an important problem within its field. }\\
        & Q2 & \textit{Addressing this challenge has the potential to significantly advance current knowledge of our research field. }\\
        & Q3 & \textit{Addressing this challenge could lead to significant improvements in practical applications.}\\
    \midrule
        \multirow{2}{*}{Urgency} & Q4 & \textit{The challenge is urgent in the context of current trends in research.} \\
        & Q5 & \textit{Addressing this challenge will prevent negative consequences that can potentially happen in practice.} \\ 
    \midrule
    Completeness & Q6 & \textit{The list is complete; they cover important challenges that hinder making visual analytics more reliable.} \\
        
    \bottomrule
    \end{tabular}
    }
    \label{tab:questions}
\end{table*}

\subsection{Expert Evaluation}

\label{sec:challeval}

We conduct an expert evaluation to verify the significance, urgency, and completeness of the challenges we identify. 
Inspired by the Deimos study \cite{lee23chi}, we recruit \revise{eight} experts in DR, data visualization, \revise{and machine learning} and conduct semi-structured interviews that ask the experts to evaluate the challenges.

\subsubsection{Study Design}
We detail how we recruit and interview participants (i.e., experts).

\paragraph{Experts and recruitment}
We recruit \revise{eight experts} who have published more than three papers related to or leverage DR in \revise{major data visualization (e.g., TVCG, VIS, PacificVis, CGF, EuroVis), HCI (e.g., CHI, IUI, TiiS), and machine learning (e.g., NeuIPS, Neurocomputing) venues}.
We recruit experts by sending them e-mails.
The demographics of the experts are depicted in \autoref{tab:demo_expert}. As of \revise{November} 2024, the participants' average number of citations and h-index is \revise{2165.9 ($\pm 3015.4$)} and \revise{15.5 ($\pm 4.9$)}, respectively, indicating their sufficient expertise in evaluating our challenges.

\paragraph{Interview protocol} 
We conduct an interview for each expert, where a single author manages all interview sessions. 
Once participants give their consent, the author explains the purpose of the research and the interview. 
The author explains that the challenges are derived from the literature review and details our workflow model and taxonomy. 
The main interview is then conducted. This part is divided into three sessions, each corresponding to the three identified challenges (\autoref{sec:challresearch}). 
During each session, the author outlines the corresponding challenge, including relevant examples and our call for the community. 
This is done by making a presentation with prepared slides. 
The experimenter then asks experts to gauge the significance and urgency of the challenge and provide their reasoning (questionnaires in \autoref{tab:questions}; Q1--Q5).
After the sessions are finished, the experimenter asks the experts to review the completeness of the set of challenges (Q6), also with reasonings.
To fully elicit the experts' insights, we do not limit the interview duration. Still, all interviews end within 50 minutes. 
An equivalent of 20 USD is paid for the compensation.

\subsubsection{Results and Discussions}
We discuss how experts evaluated our challenges. Overall, experts agree on the significance and urgency of the challenges while having mixed perspectives on the completeness.

\newcommand{\accolor}[1]{\cellcolor{appleredlight!#1}}

\begin{table*}[t]
    \centering
    \setlength{\tabcolsep}{3pt}
    \caption{\textit{The experts' evaluation on significance (Sig.), urgency (Urg.), and completeness (Co.) of our challenges.} The experts evaluate the challenges by answering the questionnaires (\autoref{tab:questions}) in Litert Scale (1: strongly disagree, 2: disagree, 3: Neutral, 4: agree, 5: strongly agree). We highlight the cell in red if the score indicates agreement, using an opacity scale where 0 represents neutral (score of 3) and 100 represents strong agreement (score of 5), with a linear gradient in between. Overall, the experts agree on the significance and urgency of our challenges while presenting mixed perspectives for completeness. 
    }
    \Description{This table presents the experts' evaluations of the significance (Sig.), urgency (Urg.), and completeness (Co.) of the challenges based on the questions outlined in the earlier table. The evaluation uses a Likert scale (1: strongly disagree to 5: strongly agree). The table highlights agreement scores using a red opacity scale, where darker red represents stronger agreement (scores closer to 5), and lighter shades represent neutrality or lower agreement (scores closer to 3). Columns: The experts (P1 to P8) evaluate the challenges across three categories: C1, C2, and C3 for each criterion. The rows also include an average (Avg.) score for each challenge across all experts. Rows: For each criterion (Significance, Urgency, and Completeness), the table breaks down responses by individual questions (Q1 to Q6). The average score for each challenge is shown at the bottom of the table. The table indicates strong overall agreement on the significance and urgency of the challenges, while there is more variability in responses related to completeness.}
    \scalebox{0.97}{
\begin{tabular}{rr|ccc|ccc|ccc|ccc|ccc|ccc|ccc|ccc||ccc}
\toprule
& & \multicolumn{3}{c|}{\textbf{P1}} & \multicolumn{3}{c|}{\textbf{P2}} & \multicolumn{3}{c|}{\textbf{P3}} & \multicolumn{3}{c|}{\textbf{P4}} & \multicolumn{3}{c|}{\textbf{P5}} & \multicolumn{3}{c|}{\textbf{P6}} & \multicolumn{3}{c|}{\textbf{P7}} & \multicolumn{3}{c||}{\textbf{P8}} & \multicolumn{3}{c}{\textbf{Avg.}}\\
\cmidrule(lr){3-5}\cmidrule(lr){6-8}\cmidrule(lr){9-11}\cmidrule(lr){12-14}\cmidrule(lr){15-17}\cmidrule(lr){18-20}\cmidrule(lr){21-23}\cmidrule(lr){24-26}\cmidrule(lr){27-29}
& & \textbf{C1} & \textbf{C2} & \textbf{C3} & \textbf{C1} & \textbf{C2} & \textbf{C3} & \textbf{C1} & \textbf{C2} & \textbf{C3} & \textbf{C1} & \textbf{C2} & \textbf{C3} & \textbf{C1} & \textbf{C2} & \textbf{C3} & \textbf{C1} & \textbf{C2} & \textbf{C3} & \textbf{C1} & \textbf{C2} & \textbf{C3} & \textbf{C1} & \textbf{C2} & \textbf{C3} & \textbf{C1} & \textbf{C2} & \textbf{C3}  \\
\midrule
\multirow{3}{*}{\textbf{Sig.}} & \textbf{Q1} & \accolor{50}4 & \accolor{50}4 & \accolor{50}4 & \accolor{100}5 & \accolor{100}5 & \accolor{100}5 & \accolor{50}4 & \accolor{50}4 & \accolor{100}5 & \accolor{50}4 & \accolor{50}4 & \accolor{50}4 & \accolor{50}4 & \accolor{50}4 & \accolor{50}4 & \accolor{100}5 & \accolor{50}4 & \accolor{100}5 & \accolor{100}5 & \accolor{100}5 & \accolor{100}5 & \accolor{100}5 & 3 & \accolor{50}4 & 4.50 \accolor{75}& 4.13 \accolor{56}& 4.50 \accolor{75}\\
& \textbf{Q2} & 2 & \accolor{50}4 & 2 & \accolor{100}5 & \accolor{100}5 & \accolor{100}5 & \accolor{50}4 & \accolor{50}4 & \accolor{50}4 & \accolor{100}5 & \accolor{50}4 & 3 & \accolor{50}4 & \accolor{100}5 & \accolor{50}4 & \accolor{100}5 & \accolor{50}4 & \accolor{100}5 & \accolor{50}4 & 3 & \accolor{100}5 & \accolor{100}5 & \accolor{50}4 & \accolor{50}4 & 4.25 \accolor{63}& 4.13 \accolor{56}& 4.00 \accolor{50}\\

& \textbf{Q3} & 3 & \accolor{100}5 & \accolor{100}5 & \accolor{100}5 & \accolor{100}5 & \accolor{100}5 & \accolor{50}4 & \accolor{50}4 & \accolor{100}5 & \accolor{50}4 & \accolor{50}4 & \accolor{100}5 & \accolor{100}5 & \accolor{100}5 & \accolor{50}4 & \accolor{100}5 & \accolor{100}5 & \accolor{100}5 & \accolor{100}5 & \accolor{100}5 & \accolor{100}5 & \accolor{100}5 & 3 & \accolor{100}5 & 4.50 \accolor{75}& 4.50 \accolor{75}& 4.88 \accolor{94}\\
\midrule
\multirow{2}{*}{\textbf{Urg.}} & \textbf{Q4} & \accolor{50}4 & \accolor{50}4 & \accolor{100}5 & \accolor{100}5 & \accolor{50}4 & \accolor{100}5 & \accolor{50}4 & \accolor{50}4 & \accolor{50}4 & \accolor{100}5 & \accolor{100}5 & \accolor{100}5 & \accolor{100}5 & 3 & \accolor{50}4 & \accolor{100}5 & 3 & \accolor{50}4 & \accolor{50}4 & \accolor{50}4 & \accolor{100}5 & \accolor{100}5 & 3 & \accolor{50}4 & 4.63 \accolor{82}& 3.75 \accolor{38}& 4.50  \accolor{75}\\

 & \textbf{Q5} & 3 & \accolor{50}4 & 2 & \accolor{100}5 & \accolor{50}4 & \accolor{100}5 & \accolor{50}4 & 3 & \accolor{50}4 & \accolor{50}4 & \accolor{50}4 & \accolor{100}5 & \accolor{100}5 & \accolor{50}4 & \accolor{50}4 & 3 & \accolor{50}4 & \accolor{50}4 & \accolor{100}5 & \accolor{50}4 & \accolor{100}5 & \accolor{50}4 & \accolor{50}4 & \accolor{50}4 & 4.13 \accolor{56}& 3.88 \accolor{44}& 4.13 \accolor{56}\\
 \midrule
 \textbf{Co.} & \textbf{Q6} &  \multicolumn{3}{c|}{\accolor{50} 4}  & \multicolumn{3}{c|}{2} & \multicolumn{3}{c|}{3} & \multicolumn{3}{c|}{\accolor{50} 4} & \multicolumn{3}{c|}{\accolor{50} 4} & \multicolumn{3}{c|}{\accolor{50} 4} & \multicolumn{3}{c|}{\accolor{50} 4} & \multicolumn{3}{c||}{2} & \multicolumn{3}{c}{\accolor{19}3.38}\\
\bottomrule
\end{tabular}
    }
    \label{tab:results}
\end{table*}

\paragraph{Significance and urgency of the challenge 1 (Lack of human-centered experiments)}
Experts generally agree with the significance and urgency of Challenge 1. As seen in \autoref{tab:results}, the average scores for all questions are greater than 4.00. Experts' reasoning also backs up the significance and urgency of this challenge. 
For example, P2 note, \textit{``The way we perceive clusters or patterns is not the way the machine computes such clusters and patterns.''}.

\paragraph{Significance and urgency of the challenge 2  (Lack of investigation on visual representations beyond 2D scatterplots)}
As with challenge 1, the evaluation scores indicate that experts overall agree with the significance of the second challenge (\autoref{tab:results}). 
Meanwhile, \revise{participants do not strongly agree on the urgency. Three participants (P1, P2, and P8)} mention that the challenge will be difficult to address as they also hardly imagine visual representations other than scatterplots, which implies that more research endeavors should be invested to resolve this challenge. 

\paragraph{Significance and urgency of the challenge 3 (Lack of libraries)}
The experts overall agree or strongly agree that this challenge is timely and significant (\autoref{tab:results}). \revise{Six} experts strongly agree with Q3, which asks about the practical impact of addressing the challenge. 
The experts also agree on the importance of building libraries from a research perspective. 
P4 especially suggests that \textit{``It will be good if we have a library in which everyone can build on top of it, just like D3.''}, reaffirming the need for the unified library that covers the entire visual analytics workflow using DR.

\paragraph{Completeness of the challenges}
The diverging scores highlight the \revise{eight} experts' differing opinions on the completeness of the challenges
(\autoref{tab:results}). For example, P5 generally agree that the main challenges were well-covered, stating, \textit{``You have hit on three really interesting areas to advance this field.''}.
On the other hand, P2 and P3\revise{, and P8} raise concerns about the completeness of the study. For example, they note that although scalability and computational efficiency do not directly improve reliability, they can indirectly do so by enabling the testing of more configurations.
The findings, overall, suggest that unknown challenges remain in the field, and we should invest more effort in systematically identifying and addressing them.

\section{Discussions}

We extend our discussion by emphasizing the practical value of our contributions and proposing future work to strengthen such aspects.

\subsection{Call for the Analysts}

\label{sec:callanalysts}

This research mainly motivates visualization researchers to make visual analytics using DR more reliable. We also detail research challenges (\autoref{sec:challanges}) that researchers can put effort into. 
We believe, however, that \analysts can also contribute to advancing visual analytics, as they are the end users of the products contributed by researchers. We provide two recommendations for analysts that can enhance the reliability of visual analytics using DR.

\paragraph{Share the analysis if possible}
We suggest \analysts deploy their analysis if there are no ethical problems, e.g., data privacy. For example, analysts can open-source their code and dataset or publicize analytic dashboards made by business intelligence platforms, e.g., Tableau or Spotfire. This will help researchers build a knowledge base for understanding visual analytics practices \cite{oppermann21tvcg}. For instance, the deployed codes and dashboards will support researchers in empirically validating and revising our workflow model (\autoref{sec:workflow}).

Moreover, researchers can leverage these deployments to address the challenges we identify (\autoref{sec:challanges}). For example, by investigating which DR techniques or evaluation metrics are widely used in visual analytics, researchers can determine which techniques or metrics should be prioritized when developing the libraries (Challenge 3). 
Deploying datasets can also benefit researchers by enabling them to include more diverse datasets in human-centered experiments, resulting in more generalizable findings (Challenge 1). 

\paragraph{Consider diverse configurations}
We also recommend \analysts consider diverse configurations.
For example, we suggest analysts test and leverage diverse DR techniques beyond $t$-SNE, UMAP, and PCA. Although not yet validated, we observe that these three techniques dominate the selection of DR methods in visual analytics, despite being unsuitable for certain tasks like cluster density estimation or global pairwise distance investigation \cite{xia22tvcg, jeon23vis, moor20icml, narayan21nature}. Using alternative techniques that better align with these tasks will enhance the reliability of their analysis. 
Additionally, if analysts more actively consider diverse configurations, it will motivate researchers to contribute new techniques that address gaps in the literature, creating a positive feedback loop for making visual analytics more reliable.

Here, a barrier to considering diverse configurations will be a \textbf{lack of operational knowledge}. 
We thus suggest \analysts refer to the \Instructor-type papers, especially the ones contributed \textbf{human-centered} and \textbf{computational experiments} to be aware of the pros and cons of diverse configurations.

\subsection{Reliability Problems Derived from Analysts}

In our survey, we find many papers that aim to make analysts more \textbf{informed}.
They provide analysts with operational knowledge of different configurations, thereby reducing the chances of analysts making suboptimal configurations.
However, though analysts are aware of reliability problems and various configurations, visual analytics can be unreliable due to human bias or errors. We discuss two cases in which awareness does not guarantee reliability.

\paragraph{Suboptimal configuration (despite awareness)}
\analysts can still make suboptimal configurations even though they are informed of corresponding operational knowledge. As discussed in \autoref{sec:challresearch}, one notable factor is the availability of actionable codes or libraries. Another factor potentially leading analysts to suboptimal configurations is their preference for the projections. For example, Morariu et al. \cite{morariu23tvcg} and Bibal and Fr\'enay \cite{bibal16arxiv} quantitatively show that separability between class labels significantly affect human preferences for the projections. Such preferences can bias analysts in setting configurations, e.g., by selecting DR techniques that emphasize cluster structure like UMAP \cite{jeon24tvcg}. Identifying factors that can potentially bias configurations and mitigate their effect will be an important future avenue to explore. For example, we can explore how visualizing distortions (\Explainer) can mitigate the biased preferences towards well-separated clusters. 

\paragraph{Inappropriate Sensemaking}
Even if analysts establish appropriate configurations, human \Sensemaking can be unreliable. 
At first, analysts are \textbf{unstable} in perceiving data patterns \cite{jeon24tvcg}, which means that their sensemaking varies by individuals and also by time. Analysts may also place excessive trust in their configurations, even for tasks that the configurations were not originally designed to support (\autoref{sec:callanalysts}).

These examples indicate that our survey does not yet fully cover all potential reliability issues in visual analytics using DR, highlighting the need for future efforts to uncover problems that remain undetected.

\subsection{Limitations and Future Work}

Our three main contributions---workflow model, taxonomy, and challenges---are theoretically grounded by literature. As future work, we successively aim to verify their validity with real \analysts. 
For example, we could crawl source codes from GitHub that involve high-dimensional data analysis using DR and evaluate how well they fit with our workflow and taxonomy. Additionally, we can conduct an interview study with industry analysts to reveal their daily challenges in practical data analysis and see how these align with our taxonomy.

We also plan to make our findings to be more actionable.
Our survey informs that using proper evaluation metrics is crucial to obtain \textbf{accurate} and \textbf{optimal} DR projections. However, as a paper, it does not recommend specific metrics to use reflecting the current intentions of analysts. 
Moving forward, we intend to develop a database that houses specific knowledge extracted from the papers we gather. 
This will facilitate the creation of an actionable framework that enables analysts to query their intention and obtain responses about the appropriate configurations, similar to the approach taken by Draco \cite{moritz19tvcg} for general visualization design.

\subsection{Call for the HCI field: Mitigating Imbalance in Research Topics}

\label{sec:mitigation}

Our meta-analysis (\autoref{sec:metaanalysis}) reveals the imbalance of the relevant research landscape: while numerous DR techniques are proposed (\Pioneer-type papers), only a few works that evaluate or interpret these techniques emerged (\Judge, \Instructor, \Explorer, \Explainer, and \Architect-type papers) (\autoref{sec:implication}). 
This imbalance aligns with the dominance of papers from the visualization and machine learning fields in our list, with only two papers originating from the HCI field. The true extent of the imbalance may be larger, as our survey started by searching for papers primarily in the visualization and HCI fields.
Such findings indicate that the HCI field should pay more attention to this issue. \revise{We call on the community to invest more research efforts in two key perspectives: \textit{System} and \textit{Guidance}.}

\paragraph{\revise{Efforts in a system perspective}}
The HCI community has already verified its capability to \revise{build new systems that} enhance the explainability and usability of machine learning techniques. For example, the community substantially contributes to the explainable AI (XAI) systems \cite{ehsan22chi, francoise21uist, gorter22chi} and also designs numerous methodologies to make AI techniques more user-friendly \cite{ha24chi, seo24chi}.
We strongly suggest the HCI community utilize such capability to make DR-based visual analytics more reliable and reduce this imbalance.
\revise{For instance, designing interactive systems that help novice analysts easily be \textbf{aware} of \textbf{inaccurcy} of \textbf{suboptimality} of DR techniques would be an interesting avenue to explore.}

\paragraph{\revise{Efforts in a guidance perspective}}
\revise{
The imbalance reveals that while visualization and machine learning fields dedicate substantial effort to advancing state-of-the-art DR techniques, their endeavors are less human- or user-centered. 
These fields seldom invest in making these techniques informed and approachable for everyday use by analysts. As a result, expert researchers and analysts are often familiar with these techniques, whereas novice analysts frequently remain unaware of their existence or applicability. To address this gap, we propose a shift in focus within the HCI community toward creating practical, user-friendly guidance. For example, interview studies can be conducted to investigate why practitioners are \textbf{uninformed} of DR techniques or evaluation metrics. Based on the insights from the studies, user-friendly guidance tailored to varying levels of expertise can be established to improve the adoption and usability of advanced methods.
}

\revise{
As a seminal step in this direction, we provide a practical guide to accompany our survey (\autoref{tab:taxonomy1}) to help analysts select the most relevant papers to read based on their expertise level and interests, thereby helping analysts improve the reliability of their visual analytics using DR.
This guide addresses a key challenge: While our survey offers a comprehensive overview of the research landscape, providing the same references to all readers is not effective because analysts have varying levels of expertise in DR. In addition, many analysts may not accurately assess their own expertise, making it difficult for them to determine the appropriate next steps in their learning journey. To address these challenges, our guide includes a six-item checklist designed to help analysts evaluate their expertise level. Based on this self-assessment checklist, we provide tailored recommendations for reading papers. Our guide can thus be interpreted as a suggestion for an optimal order for exploring the literature, enabling a more comprehensive and systematic understanding of the relevant field. We release the guide as an interactive web page to enhance readability and approachability, available at \url{https://dr-reliability.github.io/guide}.
}

\section{Conclusion}

This paper presents several contributions based on a literature review of 133 papers. This includes (1) a workflow model that explains how analysts employ DR in visual analytics, (2) a taxonomy to organize the relevant literature, (3) a meta-analysis that groups the papers at a high level, (4) a discussion of ongoing challenges validated by experts, and (5) \revise{an identification of the imbalance in research landscape and the call for the action.} Together, these contributions substantially benefit visual analytics using DR by making it more reliable. 
We expect our research to be a key reference that initiates new endeavors to enhance the reliability of visual analytics.

\begin{acks}

This work was supported by the National Research Foundation of Korea (NRF) grant funded by the Korean government (MSIT)  (No. 2023R1A2C200520911), the Institute of Information \& communications Technology Planning \& Evaluation (IITP) grant funded by the Korean government (MSIT) [NO.RS-2021-II211343, Artificial Intelligence Graduate School Program (Seoul National University)]. 
This research is also supported in part by the U.S. National Science Foundation under Grant No. IIS-2427770, U.S. National Institute of Health under Grant No. 5R01CA270454-03, and the Knut and Alice Wallenberg Foundation through Grant KAW 2019.0024.
The ICT at Seoul National University provided research facilities for this study. 
Hyeon Jeon is in part supported by Google Ph.D. Fellowship.
We thank Sungbok Shin for his valuable comments that helped improve this work. For the purpose of open access, the authors have applied a Creative Commons Attribution (CC-BY) license to any Author Accepted
Manuscript version arising from this submission.

\end{acks}
\bibliographystyle{ACM-Reference-Format}
\bibliography{ref}










\end{document}